\documentclass{aa}  

\usepackage{graphicx}
\usepackage{epstopdf}
\usepackage{amssymb}
\usepackage{amsmath}
\usepackage{multirow}

\newcommand{\bt}{\begin{turn}{90}}
\newcommand{\et}{\end{turn}}

\newcommand{\avg}[1]{\left< #1 \right>} 

\usepackage{txfonts}

\begin{document} 
\title{Statistical characterization of polychromatic absolute and differential squared visibilities obtained from AMBER/VLTI instrument
}
\titlerunning{Statistical characterization of absolute and differential squared visibilities}
\author{A. Schutz\thanks{The fellowship of A. Schutz for the present work was funded by the french ANR project POLCA (ANR-2010-BLAN-0511-02). One aim of POLCA is to elaborate dedicated algorithms for model-fitting and image reconstruction using polychromatic interferometric observations.}
          \and  M. Vannier \and D. Mary \and A. Ferrari \and F. Millour  \and R. Petrov}
   \institute{Laboratoire J.L. Lagrange, Universit{\'e} de Nice-Sophia Antipolis, CNRS UMR 7293, Observatoire de la C\^ote d'Azur, Nice, France.\\ \email{aschutz@oca.eu}             }
   \date{}

  \abstract
   {In optical interferometry, the visibility squared modulus are generally assumed to follow a Gaussian distribution and to be independent of each other. A quantitative analysis of the relevance of such assumptions is important to help improving the exploitation of existing and upcoming multi-wavelength interferometric instruments. }
   {Analyze the statistical behaviour of both the absolute and the colour-differential squared visibilities: distribution laws, correlations and cross-correlations between different baselines.}
   {We use observations of stellar calibrators obtained with AMBER instrument on VLTI in different instrumental and observing configurations, from which we extract the frame-by-frame transfer function.  
   Statistical hypotheses tests and diagnostics are then systematically applied. We also compute the same analysis after correcting the instantaneous squared visibilities from the piston and jitter chromatic effects, using a low-order fit subtraction.}
   { For both absolute and differential squared visibilities and under all instrumental and observing conditions, we find a better fit for the Student distribution than for the Gaussian, log-normal and Cauchy distributions.
We  find  and analyze clear correlation effects caused by atmospheric perturbations.
The {\it differential} squared visibilities allow to keep a larger fraction of data with respect to selected absolute squared visibilities and thus benefit from reduced temporal dispersion, while their distribution is more clearly characterized. 
}
{ 
The frame selection based on the criterion of a fixed SNR value might result in either a biased sample of frames or in a too severe selection. We suggest instead an adaptive frame selection procedure based on the stability of the modes of the observed squared visibility distributions. In addition, taking into account the correlations effects between measured squared visibilities should help improving the models used in inverse problems and thus the accuracy of model fitting and image reconstruction  results. Finally, our results indicate that re-scaled differential squared visibilities usually constitute a valuable alternative estimator of squared visibility.
}
   
\keywords{interferometry --
		calibration -- visibility --
                precision of interferometric measurements
               }

   \maketitle

\section{Introduction}
\subsection{Context and scope}

Stellar interferometers deliver data that are related  to the Fourier Transform (FT) of the intensity distribution perpendicular to the line of sight.  Ideally such interferometers are able to measure complex visibilities, which correspond to complex samples of this FT at spatial frequencies defined by the positions of the interfering telescopes or antennas, and by the observation wavelength $\lambda$.  In essence measuring moduli and phases of complex visibilities amounts to measuring contrasts and  phases of interference fringes \citep{cornwell87,labeyrie75}. 

In contrast to radio interferometric arrays however,  current optical interferometers cannot measure the phases of the complex Fourier samples because of rapid ($10$ to $20$ ms)
perturbations caused by atmospheric turbulence. Instead, they provide both their moduli (the so-called {\it{absolute visibilities}} or their squared value (the {\it{power spectrum}}) by measuring contrasts in snapshot mode with short integration times that freeze the turbulence, and a linear relationships between their phase (the {\it{phase closure}}) \citep{VisEst1984,CP1986,VisEst1994}. 

The (squared) visibility moduli are without a doubt the instrumental interferometric data that is the most used by the astronomical community. It is generally assumed that the distribution of snapshot squared visibilities follow a Gaussian distribution \citep{MIRA,meimon2005,LITPRO2008}. Moreover, in absence of systematic  processing and analysis of possible correlations, visibilities are also considered in practice as  uncorrelated. To our knowledge however, these assumptions are not established by a detailed statistical analysis, and this is the first  objective of the present paper. Obviously, the accuracy of such assumptions deeply impacts the subsequent extraction of the astrophysical information,  through  non linear least squares fits or image reconstruction.

With polychromatic optical interferometric instruments (either existing ones such as AMBER \citep{AMBER2000SPIE} or VEGA \citep{Vega2009}, or instruments in development such as MATISSE \citep{MATISSE} and GRAVITY \citep{GRAVITY}), interference fringes are obtained over several wavelengths. This allows to study possible spectral variations in the shape of the source by investigating the relative variations of the visibilities as a function of wavelength. This quantity is usually called {\it{differential visibility}}, in a similar way as for the {\it{differential phase}} which defines the interferometric phases relatively between the observed wavelengths. Differential visibilities are known to benefit from relatively lower noise with respect to the absolute ones \citep{millour2006}, and they have lead to a number of diverse astrophysical results \citep[e.g.][]{DiffVis2007_1,DiffVis2007_2,DiffVis2012}. Differential visibilities are however not always provided by interferometric reduction pipelines. When they are, they are assumed to be independent from each other both with respect to time and spectrally. Differential visibilities are consequently not often exploited in the model-fitting or image reconstruction stages. Their statistical characterization remains to be entirely done and compared to their absolute counterparts. 
  
{ In the present paper, we therefore propose to analyze fundamental statistical properties, distribution laws and correlations of the squared visibilities and of the colour-differential ones, both obtained from the AMBER instrument of the VLTI with a variety of instrumental conditions, and reduced by the standard AMBER data reduction software (DRS) "amdlib". The reasons for working on {\it squared} moduli of visibility are: {\it 1)} it is the quantity directly used within the cost functions of inverse problems, and {\it 2)} the squared-visibility estimator issued from "amdlib" is known to be less biased than the visibility amplitude estimator (see sec.\,\ref{subsec:RedProc}). }
The  studies of differential phases and closure phases are left out of the scope of this paper.


\section{Description of the data sets}

Three distinct datasets of AMBER/VLTI observations were considered:

\begin{itemize}
\item Unit Telescopes (UTs) without fringe tracker, in medium spectral resolution (february 2012)
\item Auxiliary Telescopes (ATs)  with fringe tracker, in medium spectral resolution (january 2011)
\item ATs with fringe tracker and short frame exposure, in low-spectral resolution (november 2012) 
\end{itemize}
The detailed characteristics and names of these datasets are presented in table \ref{Tab:DataSummary}. Each dataset is made of a number of exposure files on one or several\footnote{Data set AT-MR-FT includes 5 different calibrators, some of which were observed more than once along the night: HD74772, HD77020, HD91324, HD90853 and HD9249.} objects, with an overall time span covering between 0.2\,h and 5.7\,h. Exposure files contain a number of frames with an individual integration time ranging from 0.026\,s to 2.0\,s (depending  on the stellar magnitude, on the use of the fringe tracker and on the ambient conditions). The atmospheric conditions during the observations ranged from ``good'' to ``average'', with a mean seeing between 0.6 and 1.0\,arcsec and a coherence time $\tau_0$ between 2.4 and 8.4\,ms.

All observations are made using three telescopes\footnote{For the AT-LR-FT dataset, two telescopes configurations were successively used: ATs A1-K0-J3 for the first 25 observation files of that night, and ATs A1-K0-G1 for the 12 last ones.}. The order of the telescopes $T_i-T_j-T_k$ in table\,\ref{Tab:DataSummary} defines the baselines referenced hereafter: B1 is between $T_i$ and $T_j$, B2  between $T_j$ and $T_k$, and B3 between $T_k$ and $T_i$.

\begin{table}[h!] 
\caption{Data sets are made of a number of exposure files, acquired over the "Time Span" interval. Each file contains a number of individual frames, each of them exposed during a digital integration time (DIT) and separated by the "Frame sampling" interval. The medium and low spectral resolutions of AMBER are, respectively, 1500 and 35. Atmospheric seeing is given in arcsec, and coherence time $\tau_0$ in milliseconds.} 

\small
\begin{center}
\begin{tabular}{@{}l @{} c c c} 
\hline
\hline
Dataset name 			&UT-MR-NoFT 	&AT-MR-FT 		&AT-LR-FT \\   \hline
  Date 				&2011/01/16 	&2012/02/22 		&2012/10/09 \\        
  Telescopes 			&UTs 1-2-4 	&ATs D0-I1-H0 	& ATs \\        
  Objects (Kmag)		&HD15694 & various & HD197635 \\
  Kmag				& 2.5 	& 2.04<K<3.4 	& 2.8 \\ 
  Nb of files 			&2 			&40 			& 37\\       
  Time Span 			&0.2\,h 		&5.7\,h 		& 3.5\,h \\      
  Nb frames per file 		&970 		&37 			& 1000 \\ 
  DIT (s)			&0.19 		&2.0 			& 0.026\\
  Frames sampling (s)         	&0.42		&2.5 	& 	0.06 \\
  Spectral Res.			&Medium 		&Medium 		& Low \\  
  Spectral Band 		&K			&K			& K\\      
  Fringe Tracker 		&OFF 		&ON 		&ON\\        
  
\vspace{-0.7mm} Seeing\,(arcsec)   	&0.9, 1.0, 1.0 &0.6, 0.9, 1.4 & 0.5,0.6,0.8,\\  \vspace{1.0mm} 
  (min., avg., max.)  	& & 	&\\   
\vspace{-0.7mm} $\tau_0$\,(ms)  	&5.2, 8.4, 12.8 &3.3, 3.5, 3.7	& 1.9, 2.4, 3.0\\   \vspace{0.5mm}     
 (min., avg., max.)  	& & 	&\\   
  \hline
  \end{tabular} \end{center}
 \normalsize

\label{Tab:DataSummary}       \vspace{-5mm}
\end{table}

\section{Absolute visibilities}
\subsection{{Theoretical description}}
 \label{subsec:absVHistos}
The visibilities measured by an optical interferometer are affected by various random noises and biases due to perturbations from the atmosphere, 
the instrument, the electronics, and also to some extent the calibration and reduction processing applied to the data. 

For the sake of simplicity, the additive photon noise associated with the astrophysical visibility $V_\mathrm{*}$ is omitted in the following equations\footnote{For the photon-rich observations presented here, the error $\sigma_V$ due to photon noise, derived from the measured flux, is always $<0.3\%$ {\it per exposure file} (i.e. over an average of short frames as obtained generally from DRS by Amber users) well below the measured visibility fluctuations. Nevertheless, the photon noise {\it per frame} represent an error $\sigma_V$ up to $6\%$, and is therefore a significant contributor of the short-term statistical variations studied in this paper.}. 
$V_*$ is therefore considered as a deterministic quantity and  is generally unknown (except for calibration stars).
The observed squared visibility $\widehat{V}^2$ at spatial frequency $\vec{u}$, time $t$ and spectral channel $\lambda$ can then be written as  the product \citep{amdlib2007}:
\begin{eqnarray}
 \widehat{V}^2(\vec{u},\lambda,t) =  V^2_*(\vec{u},\lambda) \,V^2_\mathrm{T}(\vec{u},\lambda,t),
 \label{eq:VobsEqVstarVt} 
 \end{eqnarray}

The term $V^2_\mathrm{T}$ represents the global transfer function\footnote{The standard calibration process, where one or several calibration source(s) with known astrophysical squared visibility is observed before and/or after the science source, is often assumed to give a good estimate of the transfer function $V^2_T(\vec{u},\lambda,t)$ at the time of the science source observation. 
The propagation and correlations of errors associated with the calibration process were studied by \cite{Perrin2003}.}, whose statistics dictates that of the observed squared visibilities. 
We propose to describe the measured variations of $V^2_\mathrm{T}$ over time and over the spectral bandwidth. 
For this purpose, we use observations obtained from {\it{calibration stars}}, which are  known, stable, symmetric, and achromatic sources in the sky. For such sources, $V^2_*$ is known accurately {\it{a priori}}, so we  can assume that the estimation error on $V^2_*$ is negligible. The ratio $\widehat{V^2_T}=\widehat{V^2}/\widehat{V^2}_*\approx \widehat{V^2}/V^2_*$ then provides a relevant stochastic variable to be studied, as this variable is close to the perturbation term that will affect the squared visibility measures of science objects in routine observations.


Individual exposure frames are known to undergo frequent and/or large visibility drops. Such drops can be described using a few loss factors on the {\it{nominal interferometer's transfer function}}, say $V_{T_0}$: 

\begin{eqnarray}
	V_T = \rho(p) ~\rho(\sigma_p)~\rho(F) ~ V_{T_0},
	\label{eq:Vt}
\end{eqnarray}

where the loss factors correspond to the following effects: 

\begin{itemize}
\item the piston $p$. In \citet{AdvancedAMBER_DR_Millour2008}, the visibility loss factor $\rho(p)$ is given as a function of the ratio $p/L_c$ (where $L_c$ is the coherence length of the fringe). When replacing $L_c$ by its wavelength dependency, the visibility loss factor can be written as: 
 \begin{eqnarray}
	\rho(p) = \exp{- A \left( \frac{p}{\lambda} \right) ^2 } ,
\label{eq:rhop}
 \end{eqnarray}
where parameter $A>0$ depends on the spectral resolution and on the refraction index of the air. 
Here the value for $p$ is the average piston over the short-frame integration. When $p$ is small enough to have $A\, (p/\lambda)^2 \ll 1$ (which should be the case when a fringe tracker is used), the visibility loss varies therefore, on a first order approximation, as $\rho(p) \approx 1 - A \left( p/\lambda \right)^2$.  

\item the piston jitter $\sigma_p$, i.e. the standard deviation of the piston variations over a given time interval, also induces a loss of visibility, due to the blurring of the fringes during the frame integration. It has been shown (see Millour, PhD thesis, 2006) that its dependency with wavelength is similar to eq.\,\ref{eq:rhop}: 
 \begin{eqnarray}
	\rho(\sigma_p) = \exp{- 2 \pi^2 \left( \frac{\sigma_p}{\lambda} \right) ^2}.
	\label{eq:rhoSigmap}
 \end{eqnarray}

Therefore, a first-order development leads similarly to a quadratic expression of the loss factor as a function of $(\sigma_p/\lambda)^2$ ; this approximation is justified when $\sigma_p/\lambda$ is small and when the piston statistic is stationary over the frame integration period.

\item {the residual wavefront error (WFE) terms of order higher than 1, due to imperfect adaptive optics correction or perturbations occurring after AO, are turned into photometric variations along each beam by the spatial filtering. 
If $F_i$ notes the photometric flux (or, equivalently, the coupling ratio) on beam $i$ before its recombination, then 
the instantaneous visibility loss factor on baseline $B_{ij}$ would be: 
 \begin{eqnarray}
	\rho(F_i,F_j) = \frac{2 \sqrt{F_i F_j}}{F_i+F_j}
	\label{eq:rhoF}
 \end{eqnarray}
{ On Amber, the beam unbalance is monitored using dedicated photometric channels, and the consequent visibility loss is largely corrected by the amdlib DRS. Nevertheless, there remains a residual visibility drop because of fast (i.e. faster than the frame integration time) variations in the beam ratio due to wavefront errors, such as high-frequency vibrations of the telescopes. To our knowledge, this residual loss within each frame cannot be corrected for. Also, this effect is probably chromatic (due to the differential atmospheric refraction affecting the beam injected in the spatial filter, and the dependency of the spatial filter refractive index with wavelength), but we have no analytical law to describe this chromatic behaviour.}
}
 
\end{itemize}

These factors depend of course on the atmospheric turbulence statistics (seeing and coherence time being the relevant parameters), on the air mass and on the setup and stability of the instrument as a whole (e.g. telescope vibrations, quality of the coherencing and cophasing by the fringe tracker, \dots). Although it is difficult to attempt describing a very general and accurate statistical perturbation model,  we shall see in the next section that this  model describes simple perturbation  behaviour that can be exploited in the reduction/calibration process.


\subsection{Reduction process}
\label{subsec:RedProc}
The observations of calibration stars were first reduced with the standard AMBER Data Reduction Software "amdlib" \citep{amdlib2007, amdlib2010}, without binning the individual frames. {  It is known that the squared visibilities estimator provided by amdlib (just as for other ABCD-based\footnote{The "ABCD" method consists in estimating the fringe contrast and phase from a sample of four measurements per fringe period. See \citet{Colavita99} for more details.} algorithms) is less biased than the estimator of complex visibility modulus. Therefore we hereafter work exclusively on squared visibility ($V^2$) data, even though we may omit occasionally, for a lighter readability and when the context is non-ambiguous, the adjective "squared". Note also that a transformation from the $V^2$ estimator to a visibility modulus estimate would not be as trivial as taking the square root of $V^2$, as in some cases the squared estimator can get negative values due to a bias effect. Following eq.\,\ref{eq:VobsEqVstarVt}, the reduced $\widehat{V^2}$ is then divided by the theoretical squared visibilities $\widehat{V^2_*}\approx V^2_*$ of the corresponding calibration star\footnote{ Their diameter, associated errors and expected squared visibilities  can be found using SearchCal (the JMMC Evolutive Search Calibrator Tool), at URL \url{http://www.jmmc.fr/searchcal_page.htm}.}. }

 
The datasets to be analyzed are thus composed of series of calibrated squared visibilities which represent estimates of the transfer function, with the time sampling of the individual frames. 

Whereas an instantaneous drop of visibility may have several causes which are hardly disentangled, the effect of piston parameters on the dependency of squared visibility with wavelength appears clearly in eq.\,\ref{eq:rhop} and \ref{eq:rhoSigmap}. These two terms can be combined as a global factor. 
As an example we can rewrite eq. \ref{eq:rhoSigmap} as:
\begin{eqnarray}
\rho(\sigma_p,\lambda_n)\approx C(\sigma_p,\lambda_m)\,\left( \left(\Delta \lambda_n\right)^2 + a\,\Delta \lambda_n \right)
\end{eqnarray}

 where $\Delta \lambda_n= \lambda_n-\lambda_m$, with $\lambda_n$ and $\lambda_m$ being respectively the $n^{\rm th}$ wavelength and the average wavelength in the considered waveband.

\begin{figure}[!ht] 
\includegraphics[width=.5\textwidth]{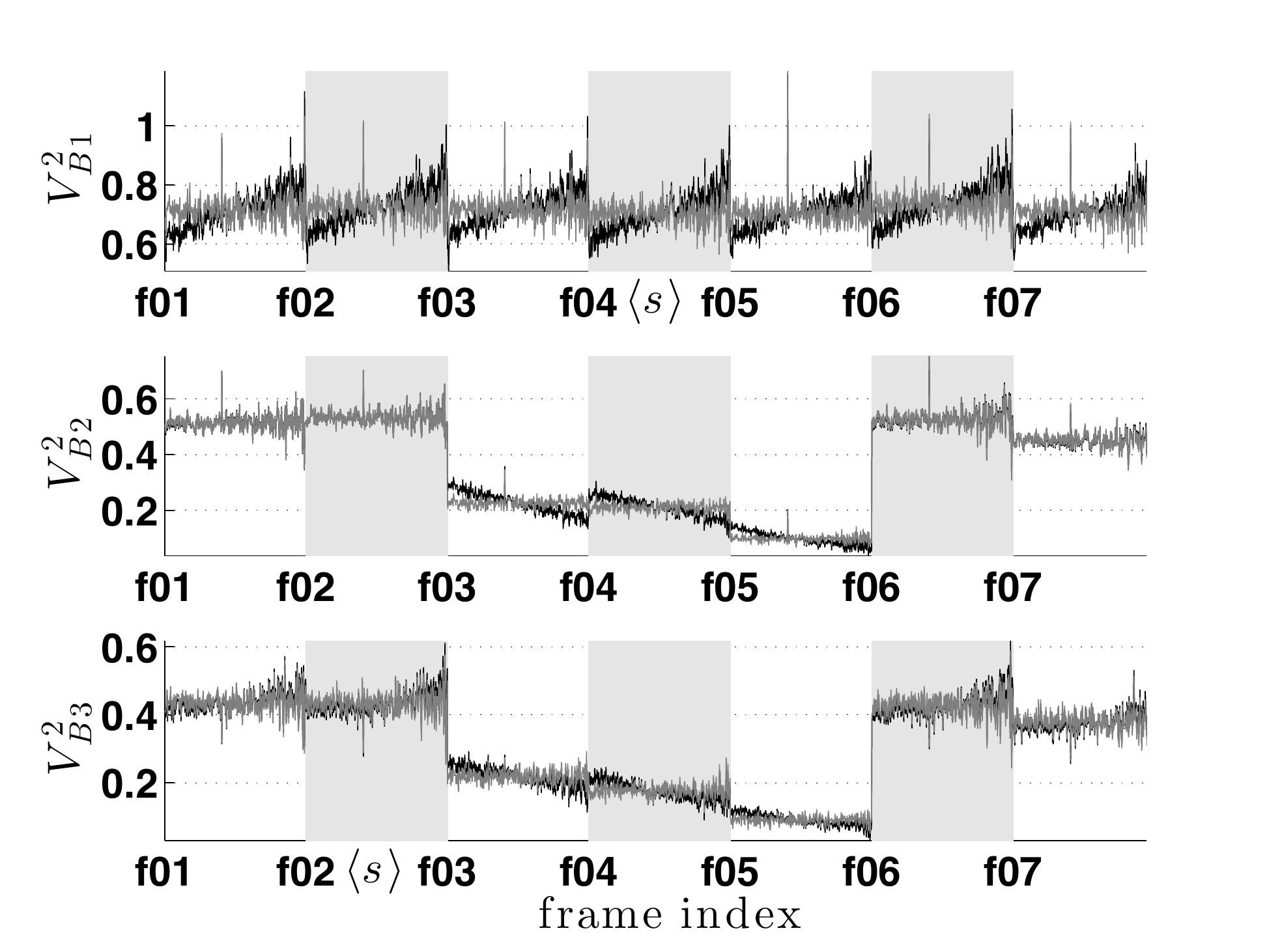}  	
\caption{Sample, from dataset AT-MR-FT, of squared visibilities before (black line) and after (grey) subtraction of a second order polynomial fit representing the effect of the piston and jitter over the spectral dimensions, for a few successive frames (f$01$ to f$07$, indicated by the alternation of background color) and for the three baselines. Each frame displays about $500$ spectral channels.The average level of visibility within each frame is conserved. The $\avg{s}$ marker on abscissa axis indicates a frame included in the "best $20\%$" selection.}
\label{Fig:AbsVisFit} 
\end{figure}

Subtracting the piston effect from the visibilities issued from science object observations is usually not done, as it would either require an accurate knowledge of $p$ and $\sigma_p$ for each frame, or a low-order fit along wavelengths in order to estimate its amplitude, following eq.\,\ref{eq:rhop} and \ref{eq:rhoSigmap}. In the latter case, removing that fitted trend would indeed also suppress some actual astrophysical signature included in the squared visibilities. 
But it can be used on calibrator observation, for the purpose of this study: our "piston-fit corrected" squared visibilities are obtain after subtracting a second order polynomial fit over the spectral dimension for each frame, without changing its average squared visibility. Such post-correction of piston effects is illustrated in figure\,\ref{Fig:AbsVisFit}, and will be used hereafter together (and in comparison) with the squared visibility data reduced in the standard way (i.e. without a correction for the piston and jitter chromatic effect).

\subsection{Distribution of squared visibilities as a function of frame selection} 
\label{subsec:absVHistos}

As presented in table \ref{Tab:DataSummary}, observation data are made of a number of short-exposure frames, each one containing integrated fringes dispersed along the spectral dimension. Because the quality of each frame depends greatly on the instantaneous turbulence conditions, the data reduction process allows to select the frames which should actually be used for estimating $V^2$, whereas the others have to be discarded.
By analyzing the distribution of the reduced squared visibilities, for all frames and spectral channels, we briefly address the question of the criterion and threshold to be used for that selection, and its impact on the resulting statistics of the visibilities: according to which criterion should individual frames be declared ``useless" and be filtered out ? For AMBER data using the standard DRS \citep{amdlib2010}, the default (and advised) criterion is the ``fringe SNR", estimated from the residuals of the reduced fringe fitted by an internal calibration fringe frame, and the default threshold is set to the best 20\% of the total number of frames, according to that criterion.  The user can otherwise choose the selection criterion\footnote{The other possible selection criteria of the AMBER DRS are the flux balance and the piston on each baseline.} and its associated threshold (either absolute or given as a fraction of the total frames).

\begin{figure}[hb!] 
\begin{center}
\includegraphics[width=.5\textwidth]{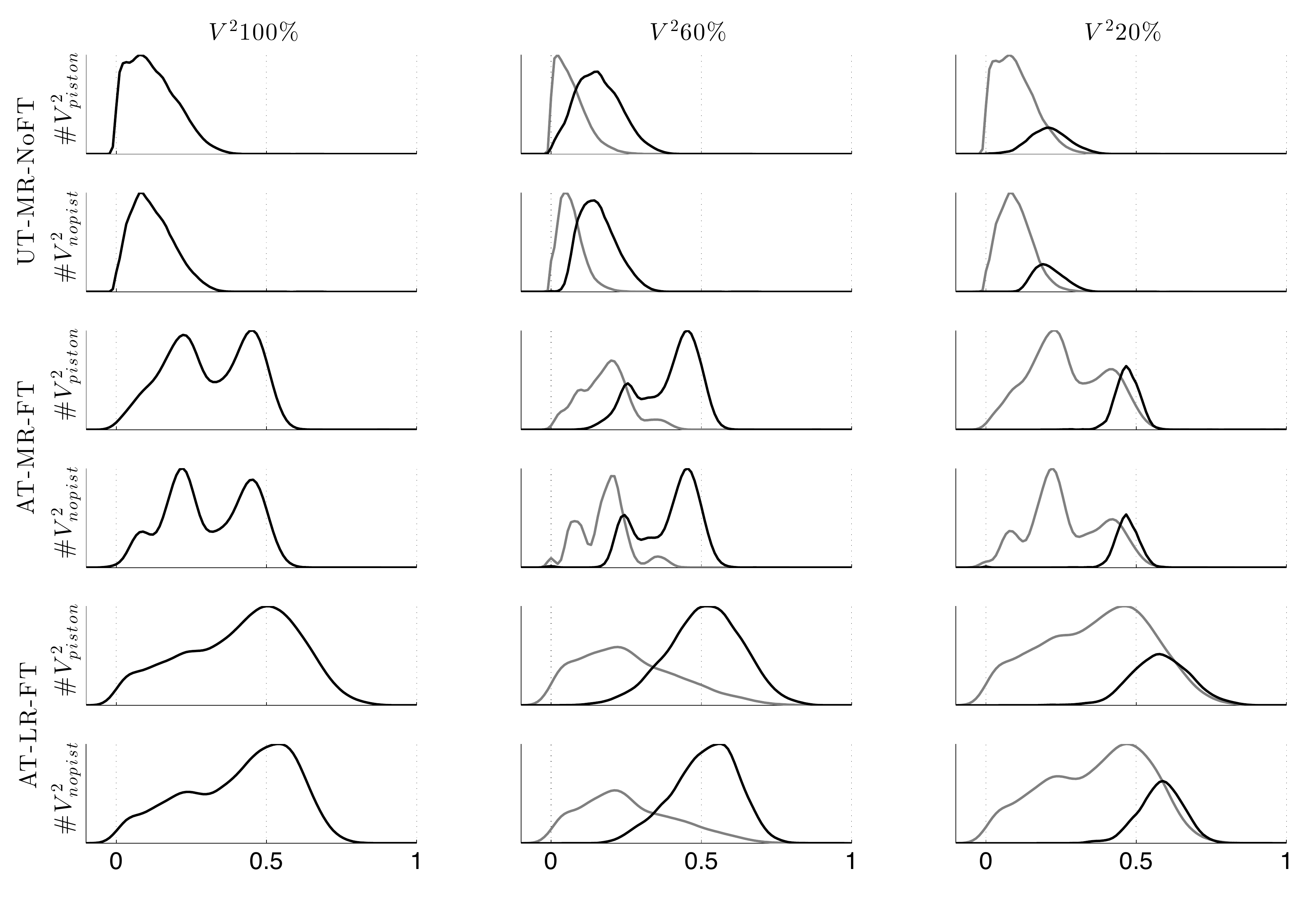}  	
\caption{Examples of histograms of $\widehat{V^2_T}$ for, from top to bottom, data sets UT-MR-NoFT, AT-MR-FT and AT-LR-FT.
The columns correspond to different threshold levels for the selection of the frames (from left to right: no selection, 60\% and 20\% of the total number of frames). 
For each dataset and selection rates, histograms are given with ($V^2_\mathrm{piston}$) and without ($V^2_\mathrm{nopist}$) correction by a quadratic fit.
The grey thin curves in solid line represent the unselected part of the data.
}
\label{Fig:HistVis}  
\end{center}
\end{figure}

Figure\,\ref{Fig:HistVis} shows some examples of squared visibilities histograms (solid black lines, one exposure file for each dataset) at three different levels of fringe SNR selections and with/without the correction of the piston effect. When no selection at all is made (left column), the  occurrence of frames with a visibility loss induces  distributions which are wide, and possibly asymmetric and multimodal. The data contain over-represented low-visibility components, appearing either as an increased left-side tail, or as one or several distinct modes. This latter case appears for observations with the fringe tracker, and probably corresponds to a degraded (offset by one or several fringes) or simply lost tracking\footnote{Making a more precise diagnosis in this case would require to compare the timing and amplitudes of the visibility losses with AMBER's fringe-tracker (FINITO) records, which were not available to us for the considered observations.}.

Making a selection of "best frames" (according to the criterion presented above) certainly regularizes the distributions, which look closer to  bell-shaped one (see the  second and third columns for two different threshold values). However, this does not seem to allow an unambiguous statistical characterization, from the following considerations:\\

 \begin{enumerate}[1)]
\item The mean and standard deviation of the distributions are dependent on the level of frame selection. They both vary with the increasing selection threshold level (the mean increases and the standard deviation decreases) up to a level where they stabilize. 
The distribution of the observed visibilities will be the focus of section \,\ref{subsec:GofStats}.

This means that unless an appropriately severe selection of frames is made, the resulting sample mean used to estimate $V^2$ may yield very different results. 
On the other hand, throwing away more measurements leads to increase the variance of the estimation error.

A similar analysis was also tried with other criteria (photometry balance, instantaneous piston,\dots) with the same threshold dependent behaviour.\\

\item  The threshold level at which values of the mean and standard deviation stabilize varies greatly with the baseline and the ambient conditions. 
For instance, in some cases a tolerant selection of 60\%  gives a distribution which is very similar to that obtained with a conservative selection level of 20\%. 
In such cases, using a too high SNR level means that 40\% of useful  data are ignored, although they are available and could be used to better estimate $V^2$. 
In other cases, the  60\% criterion is clearly too tolerant and leads to include in the estimation process data points that may convey more uncertainty than real information about $V^2$. \\

 \item The temporal behaviour of the spectral content is illustrated in figure \ref{Fig:AbsVisFit} (and more in figure \ref{Fig:HistDeltaVNoSelecUT}) where the selected frames at 20\% are denoted with a $\avg{s}$ on the abscissa. As mentioned before, the selection uses a SNR criterium to reject a part of the frames. 
The selected frames have in particular a high mean value and are only slightly affected by the piston effect. It appears, however, that some rejected frames clearly exhibit the same characteristics.

\end{enumerate}

This analysis confirms that from a set of observed short-frames, it is  difficult to derive automatically the best estimator of the visibility and a reliable related uncertainty (and 
this justifies the usefulness of having  several criteria available  for frame selection).
But this also suggests  that the frame selection procedure
 could be adaptive  instead of  using a fixed criterion such as SNR  for instance. The SNR selection  threshold  could be  increased progressively  until the distribution looks stable enough in mean for instance.
The mode of the observed distribution (instead of the sample mean) also appears as an interesting estimator to be investigated\footnote{Note that the relative stability of the mode with respect to the mean was also exploited in nulling interferometry \citep{Mennesson}}. These points deserve a detailed study and are left outside the scope of the present paper.

The correction by a quadratic fit compresses, for each frame, the squared visibility around the mean, reduces the dispersion and makes the different modes more distinguishable on the total distribution (cf figure \ref{Fig:HistVis} showing histograms for all frames and wavelengths). 
This correction mainly affects the analysis of individual frames. The dispersion reduction induced by the correction attenuates the asymmetry, or at least gives a distribution easier to analyze.
\section{Differential visibilities}
\label{secdifv}
\subsection{Construction and associated distributions}
\label{sec:diffVis}

In a general way, the colour-differential visibility $\Delta V$ is the ratio between the modulus of visibilities at the current spectral channel $\lambda$ and at a reference channel. 
The practical choice of the reference channel may vary, depending on the considered instrumental setup: it may be made of a single reference channel, or derive from a set of several spectral 
channels over which the squared visibility will be averaged. 
a central spectral feature, such as an emission or absorption line) or variable depending 
on the current channel $\lambda$. 
For AMBER data, the standard choice \citep{AMBER_Diff} is
to consider for the reference the average from all $n_\lambda$ observed spectral channels except the current one.
{ 
In the current study, the differential visibility is computed from the same estimator $V^2$ (introduced in sec\,\ref{subsec:RedProc}) used for the absolute ones. In order to allow the comparison between these quantities, we will therefore work on differential {\it squared} visibilities. According to the AMBER convention for the reference channel, these are defined,  for a current exposure frame $i$ and a given baseline, as:

 \begin{eqnarray}
\Delta V^2(i,\lambda) &=& \frac{V^2(i,\lambda)}{V^2_{\mathrm{ref}}(i)}~,\; V^2_{\mathrm{ref}}(i)= \frac{1}{n_\lambda-1} \sum_{\substack{n=1 \\ \lambda_n \ne \lambda}} ^{n_\lambda} V^2(i,\lambda_n)  \label{eq:DeltaV}
\label{defDV2}
 \end{eqnarray}


The equation above can be applied at all baselines, frames and wavelengths. We do not perform here a frame selection based on an external criterion, but only the initial filtering of bad flagged data, and an additional filtering process (see appendix \ref{ann:diffvis}) to filter out some odd points (usually less than 5\% of the total data) with unexpectedly high or low values. 


From eq.\,\ref{defDV2} it can be seen that the reference squared visibility has a mean value over wavelengths close to the empirical mean value of $V^2$ when $n_\lambda$ is large. The resulting differential squared visibility has therefore a mean value close to one, and the corresponding statistical dispersions over wavelengths and over time are also increased by a factor $1/V^2_{\rm ref}$ with respect to the absolute ones. 
}

\begin{figure}[ht!] 
\begin{center}
\includegraphics[width=.5\textwidth]{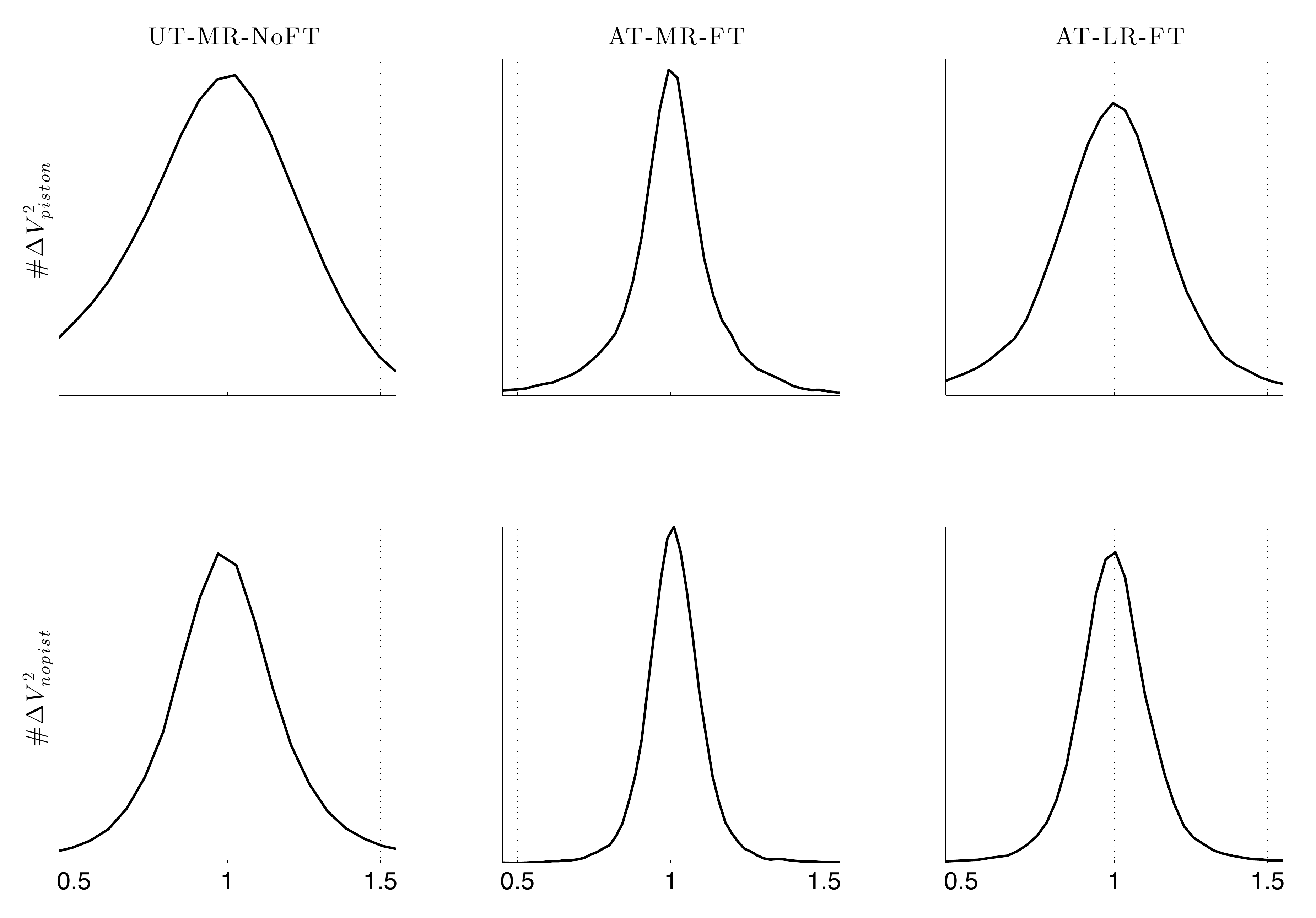}  	
\caption{Examples of differential squared visibilities histograms for the same datasets as in the left-hand column of figure \ref{Fig:HistVis} (no frame selection).}
\label{Fig:HistDiffUTaAT}
\end{center} \vspace{-7mm}
\end{figure}

Some examples of histograms appear in figure \ref{Fig:HistDiffUTaAT}. 
When compared to the non-differential squared visibility histograms with the same level of selection (left-hand column of figure \ref{Fig:HistVis}), they indicate that differential visibilities are less scattered, more symmetric and regular (thus easier to analyze) than their absolute counterparts.
 

The explanation is that the instantaneous drops which affect the absolute visibilities are, on a first order, mostly achromatic: for a given frame and baseline, they induce a loss of visibility globally over the spectral band. On a first order approximation, the differential visibility is insensitive to these visibility drops thanks to the normalization by the reference channel. On a second order only, the dependency of piston and jitter imply a variation of the observed visibility with wavelengths (eq.\,\ref{eq:rhop} and \ref{eq:rhoSigmap}), both on absolute or differential quantity. 

Note that the correction of visibility drop in $\Delta V^2$ also applies to the frames suffering from an apparent loss (or poor quality) of the fringe tracking, here in the case of AT-MR-FT observations: the data corresponding to "secondary modes" located left of the highest-visibility mode (e.g. middle-column plots in figure \ref{Fig:HistVis}) are now integrated in the centered distribution.


\subsection{Rescaled differential squared visibilities as an estimator of the squared visibilities}

Whereas we presented in the previous section $\Delta V^2$ as being normalized by the reference channel, and therefore centered close one,
 the observer still needs squared visibilities correctly scaled to the size of the source. A way to obtain this from differential squared visibilities, is to rescale them to an average level $V^2_{\rm m}$, correctly estimated from the absolute ones. For a given observation file and baseline, the rescaled differential squared visibility will then simply be: $\Delta V^2_{\rm R}(i,\lambda) = V^2_{\rm m}\,\Delta V^2(i,\lambda)$\;, for any frame $i$. 

Although the actual criterion and threshold for getting that selection remains an open subject of discussion, we nevertheless fixed these parameters by taking the default values mentioned previously (``fringe SNR'' criterion, with a best $20\%$ frames selection threshold), in order to make a comparison between the absolute and differential quantities, at a same scale.
  The averaging for computing $V^2_{\rm m}$ should be both spectral and temporal. We note $\bar{\mu} _{v^2}(i)=(1 \slash n_\lambda) \sum_{n}V^2(i,\lambda_n) $ the squared visibility at frame $i$, averaged over the spectral channels. $V^2_{\rm m}$ is the average value of $\bar{\mu} _{v^2}(i)$ over the set $\mathcal{I}^s$ of selected frames: $V^2_{\rm m} = (1 \slash n_{\#(\mathcal{I}^s)}) \sum_\mathrm{j \in \mathcal{I}^s}\bar{\mu} _{v^2}(j)$ where $n_{\#(\mathcal{I}^s)}$ stands for the number of selected frames. 

\begin{figure*}[!hp] 
\includegraphics[width=\textwidth]{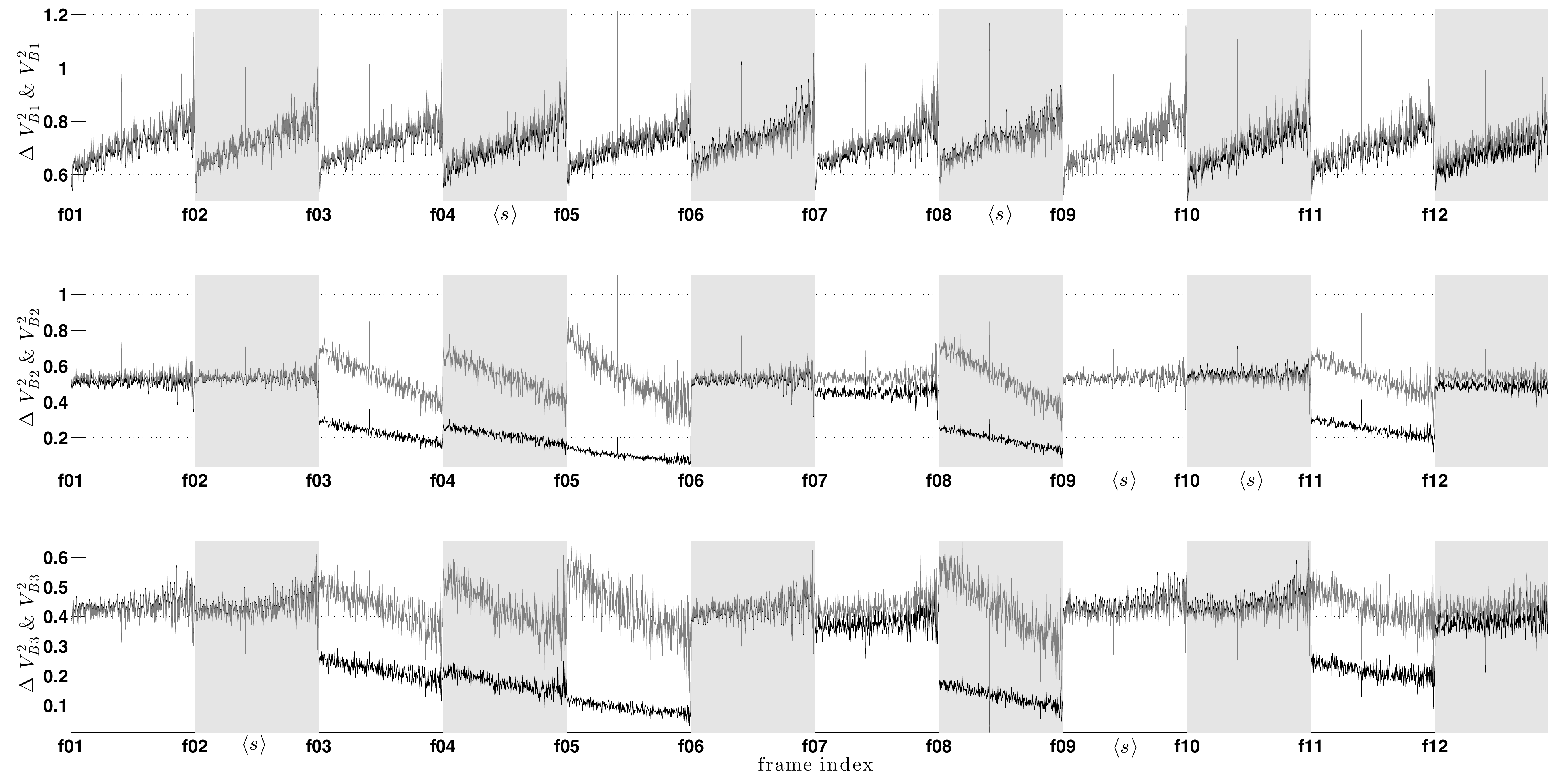}  	
\caption{Squared visibilities $V^2$ (black) and rescaled differential squared visibilities $\Delta V^2_{\rm R}$ (grey) at all three baselines, for a small sample of successive frames (indicated by the frame index and the alternation of background color) from data set AT-MR-FT, and at all wavelengths within each frame. Both quantities are superimposed for frames and baselines where the visibility is close to $V^2_{\rm m}$, the estimated average squared visibility which is used to rescale $\Delta V^2$. The $\avg{s}$ marker on abscissa axis indicates a frame included in the "best $20\%$" selection.}
\label{Fig:HistDeltaVNoSelecUT} 
\begin{center}
\includegraphics[width=\textwidth]{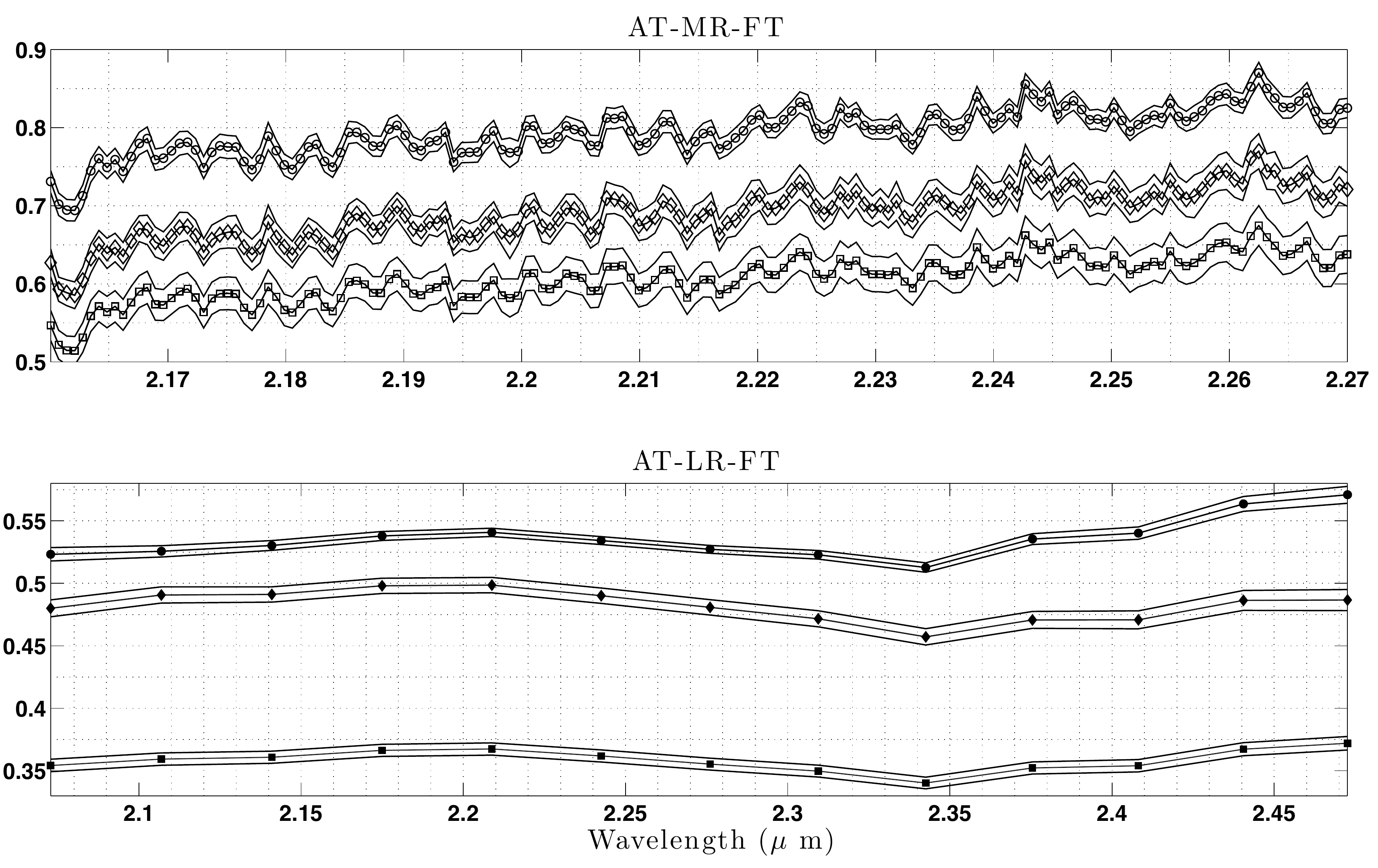}
\caption{Average values of squared visibilities and rescaled differential squared visibilities, together with their standard deviations per file, for Baseline 1 of our datasets AT-MR-FT {\it (top)} and AT-LR-FR {\it (bottom)}. squared visibilities are plotted for frame selection levels of 100\% and 20\% (square and diamond symbols, respectively). Rescaled differential squared visibility $\Delta V^2_{\rm R}$ (circle) are shifted for more clarity by vertical offsets of 0.1, and were obtained using almost 100\% of the frames (see sec.\,\ref{sec:diffVis}). In each case, the lines above and below the average values of squared visibility indicate the values of its standard deviation $\bar{\sigma}_t$ over one observation file, which are consistent with $\bar{\sigma_t}$ computed over the whole set of files, as shown in figure \,\ref{Fig:SigmaLR} and in table\,\ref{Tab:SigmaLambdRawResidual}. This figure is an illustration of the fact that the differential squared visibility has a very similar shape as the absolute ones, but with a lower statistical dispersion.}

\label{Fig:RmsAvgAT}
\end{center}  
\end{figure*}

In figure \ref{Fig:HistDeltaVNoSelecUT} (a small sample from AT-MR-FT data), both quantities are represented linearly, for a series of frames and at all wavelengths. Another illustration is shown in figure \,\ref{Fig:RmsAvgAT}, where the frame-averaged levels of $V^2$ at 20\% ($V^2_{20\%}$) selection level and of $\Delta V^2_{\rm R}$ (which is here offset vertically for more clarity) are identical by construction, and their global shape along wavelength is very similar. But they differ by some weak features and by a small slope: $\Delta V^2_{\rm R}$ is time-averaged over a sample of frames much larger than the one used for $V^2_{20\%}$, and the two samples do not necessarily have the same average piston, which determines the slope of the resulting squared visibilities. 



\section{Statistical Results} 
\label{Statres}
\subsection{Compared dispersions}

We present and compare hereafter the global statistical results, summarized in table\,\ref{Tab:SigmaLambdRawResidual}, of different estimators of squared visibility: the empirical mean of squared visibilities\footnote{All averaging operations on (squared) visibilities are in fact averages on the (squared) coherent flux. In practice, averaging directly the visibilities gives the same numerical results (down to $\approx 0.1\%$), due to the stability of the photometric measurements in our datasets} at frame selection levels of $100\%$ and $20\%$ (hereafter called $V^2_{100\%}$ and $V^2_{20\%}$), and the differential squared visibility rescaled at $V^2_{\rm m}$ ($\Delta V^2_{\rm R}$) as explained previously. All the quantities in table\,\ref{Tab:SigmaLambdRawResidual} are computed separately for each bases and exposure file, and then averaged over the exposures of each dataset.

The first and second columns, $V^2_{\rm m}$ and $\bar{\sigma}( \bar{\mu}_{v^2})$, indicate respectively the average squared visibility level and the standard deviation of $\bar{\mu}_{v^2}(i)$ along the selected frames of an exposure file. We do not discuss in this paper the longer-term variations of the transfer function, i.e the variations of $V^2_{\rm m}$ between the exposure files. As expected from the discussion in sec.\,\ref{subsec:absVHistos}, $V^2_{\rm m}$ is lower, and $\bar{\sigma}( \bar{\mu}_{v^2})$ is larger, for $V^2_{100\%}$ than for $V^2_{20\%}$. 

\begin{table}[ht!]   \vspace{0mm}
\caption{Standard deviations of the squared visibilities ($V^2_{100\%}$: without frame selection and $V^2_{20\%}$: with 20\% frames selection) and of the rescaled differential ones ($\Delta V^2_{\rm R}$), for our three datasets. If we let $\bar{\mu}_{v^2}(i)$ be  the squared visibility at frame $i$ averaged over wavelengths, then $V^2_{\rm m}$ refers to the average of that quantity over an exposure file,  and $\bar{\sigma}( \bar{\mu}_{v^2})$ its standard deviation, per frame. Columns with $\bar{\sigma}_t$ refer to the standard deviations of squared visibilities over time, per exposure file (therefore including a $1/\sqrt{N_{\rm fr}}$ factor), computed and then averaged for all wavelengths. Columns with $\bar{\sigma}_\lambda$ contain the standard deviations over wavelengths, computed from the average visibility of the frames in the considered selection. In addition, the standard deviations $\bar{\sigma}_t$ and $\bar{\sigma}_\lambda$ are also given for the considered squared visibilities corrected from the piston trend ($V^2_{\rm no pist.}$), i.e. after subtraction of a second order polynomial fit. All results are averaged over the various exposure files and baselines of a given dataset.}
\small
\begin{tabular}{c|c|c|c|c|c|c|}

\cline{2-7}

 $\%$ & $V^2_{\rm m}$ &$\bar{\sigma}( \bar{\mu}_{v^2})$ & $\bar{\sigma}_t(V^2_{})$& $\bar{\sigma}_t(V^2_{\rm no pist})$&$\bar{\sigma}_\lambda(V^2_{})$& $\bar{\sigma}_\lambda(V^2_{\rm no pist})$ \\ [1.1ex] \cline{2-7}

 \multicolumn{7}{c}{ UT-MR-NoFT } \\ \hline
 \multicolumn{1}{|c|}{$V^2_{100\%}$}               	 & 13 	 & 6.5 	  	 & 0.25 	 & 0.22 	 & 1.8  	 & 0.53	  	\\ \hline
 \multicolumn{1}{|c|}{$V^2_{20\%}$}               	 & 22 	 & 4.2 	  	 & 0.44 	 & 0.36 	 & 3.1  	 & 0.86	  	\\ \hline
 \multicolumn{1}{|c|}{$\Delta V^2_{100\%}$}               	 & 22 	 & 0.014 	  	 & 0.3 	 & 0.17 	 & 3.1  	 & 0.92	  	\\ \hline
 \multicolumn{7}{c}{ AT-MR-FT } \\  \hline
 \multicolumn{1}{|c|}{$V^2_{100\%}$}               	 & 43 	 & 14 	  	 & 2.5 	 & 2.5 	 & 4.7  	 & 4.1	  	\\ \hline
 \multicolumn{1}{|c|}{$V^2_{20\%}$}               	 & 55 	 & 2.1 	  	 & 2.3 	 & 2.3 	 & 6.1  	 & 5.6	  	\\ \hline
 \multicolumn{1}{|c|}{$\Delta V^2_{100\%}$}               	 & 55 	 & 0.0086 	  	 & 1.7 	 & 1.4 	 & 06  	 & 5.4	  	\\ \hline

\multicolumn{7}{c}{ AT-LR-FT } \\ \hline
 \multicolumn{1}{|c|}{$V^2_{100\%}$}               	 & 36 	 & 14 	  	 & 0.5 	 & 0.48 	 & 3.2  	 & 0.81	  	\\ \hline
 \multicolumn{1}{|c|}{$V^2_{20\%}$}               	 & 49 	 & 6.3 	  	 & 0.6 	 & 0.55 	 & 4.6  	 & 1.1	  	\\ \hline
 \multicolumn{1}{|c|}{$\Delta V^2_{100\%}$}               	 & 50 	 & 0.57 	  	 & 0.43 	 & 0.29 	 & 4.9  	 & 1.2	  	\\ \hline

%
%
\end{tabular}

\label{Tab:SigmaLambdRawResidual}  
\end{table} \vspace{0mm}

On the other hand, the average squared visibilities of $\Delta V^2_{\rm R}$ and $V^2_{20\%}$ are almost identical\footnote{Some small ($< 1\%$) difference between $\Delta V^2_{\rm m, 100\%}$ and  $V^2_{\rm m, 20\%}$ (the total empirical mean for the differential squared visibility at $100\%$ and the squared visibility at $20\%$ respectively) appear and can be explained by the fact that the average of the differential squared visibility before rescaling is close, but not equal, to 1, since it uses a reference squared visibility distinct from the average squared visibility of a given frame (eq.\,\ref{eq:DeltaV}).}
and $\bar{\sigma}({\Delta V^2_{\rm R}})$ is almost null: this simply follows the definition of $\Delta V^2_{\rm R}$ in sec.\,\ref{subsec:absVHistos}, whose average on each frame is set at $V^2_{\rm m}$, in other word the differential squared visibilities did not suffer of visibility loss.

The results given hereafter compare the dispersions of both absolute and differential squared visibilities along: 
{\it 1)} the time dimension, considering all the selected frames within an exposure, the spectral channels being considered separately and eventually averaged.
{\it 2)} the wavelengths, computed after a previous averaging of the squared visibilities from the selected frames, within an exposure files. 

In order to allow a comparison of both the "flatness" over wavelength and the stability over time of the squared visibilities with and without the piston influence mentioned in sec.\,\ref{subsec:RedProc}, the same quantities were also studied after subtraction of a second order polynomial fit in each frame, while conserving their average value (see figure \,\ref{Fig:AbsVisFit}). These  "flattened squared visibilities" statistics are referred hereafter as $V^2_{\rm no pist.}$.

Along the time dimension, the standard deviation \emph{per exposure file} $\sigma_t$ can be estimated as the empirical standard deviation per frame divided by the square root of the number of selected frames $N_{\rm fr}$ within that considered data sample.

\begin{figure}[ht!] 
\includegraphics[width=.5\textwidth]{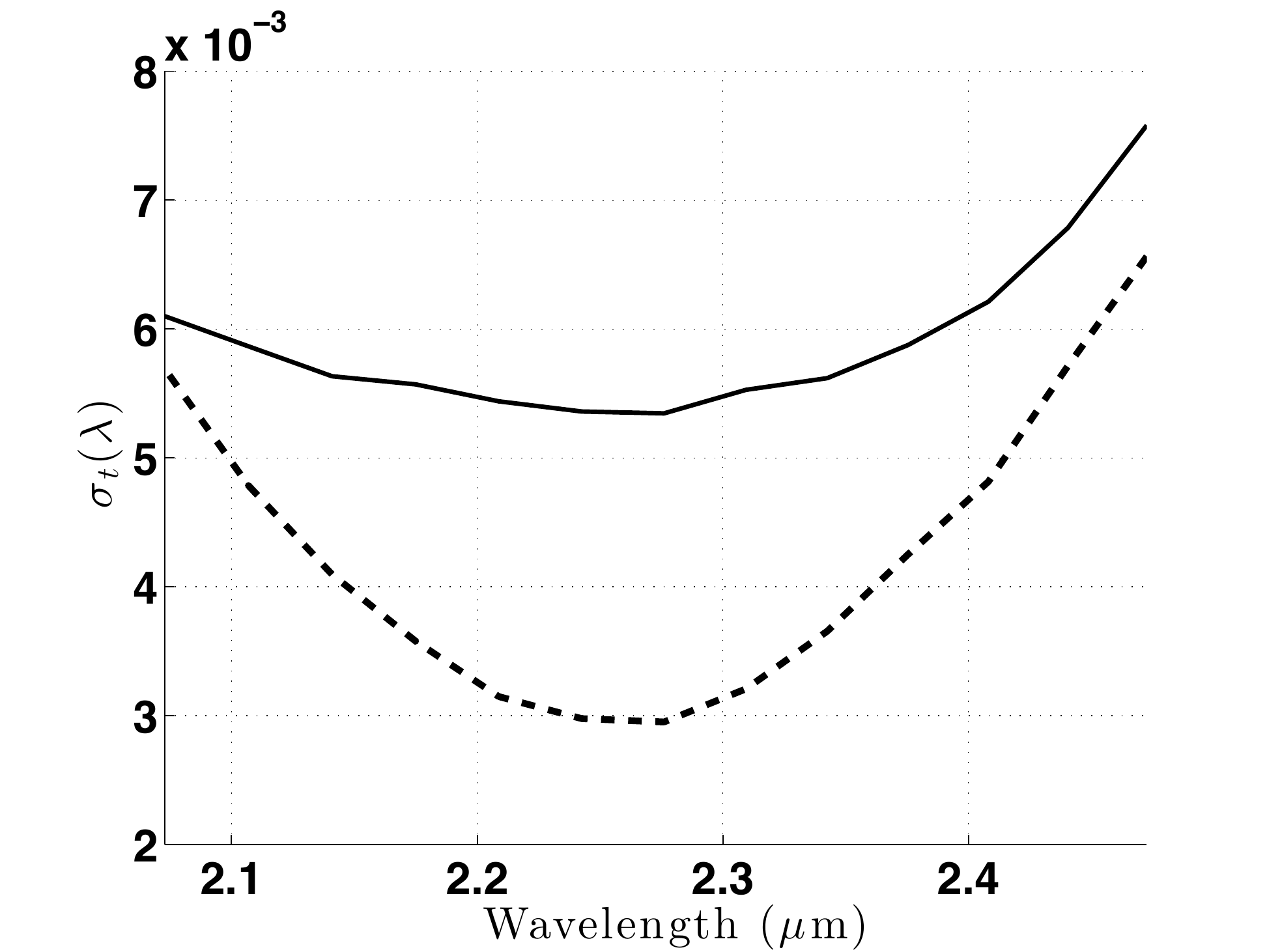}  	
\caption{Standard deviation over time $\bar{\sigma}_t$ of squared visibilities (20\% frames selection, solid line) and of rescaled differential squared visibilities (dashed line), for each wavelength, \emph{per exposure file} and for dataset AT-LR-FT. $\bar{\sigma}_t$ is averaged over the three baselines and over the different exposure files of the dataset. The lower statistical dispersion of the differential squared visibility compared to the absolute one appears clearly in this figure. The mean values of $\bar{\sigma}_t$ over wavelengths are presented in table\,\ref{Tab:SigmaLambdRawResidual} for all the datasets.}
\label{Fig:SigmaLR} 
\end{figure}  

The error bars surrounding the curves in figure \ref{Fig:RmsAvgAT} and the figure \ref{Fig:SigmaLR} (for the AT-LR-FT case), show $\bar{\sigma}_t$ as a function of wavelength, for $V^2_{20\%}$ and $\Delta V^2_{\rm R}$, without a correction of the piston effect. These numbers are averaged over wavelengths and baseline and completed by the results for the piston-fitted quantity in table\,\ref{Tab:SigmaLambdRawResidual} (resp. columns $\bar{\sigma}_{t}(V^2)$ and $\bar{\sigma}_{t}(V^2_{\rm no pist})$). 

It appears that, for all the datasets, the standard deviations over time of the rescaled differential squared visibility $\Delta V^2_{\rm R}$ are lower 
  than those of the absolute one, 
  either with or without frame selection. The improvement factor on $\bar{\sigma}_t$ is about 25\%, with respect to the standard $V^2_{20\%}$ estimator. The frames sample of $\Delta V^2_{\rm R}$ include relatively more frames of lower quality and more scattered squared visibilities (and thus a higher standard deviation \emph{per frame}), but since it is also a much larger sample (almost 100\% of frames vs. 20\%), the standard deviation \emph{per file} is finally improved ; in other terms, the $1/\sqrt{N_{\rm fr}}$ factor improves the statistical error in favor of the non-selective $\Delta V^2_{\rm R}$ estimator. Unsurprisingly, the possibility to subtract a fitted piston effect also induces a significant improvement (between 10\% and 30\%).

As for the standard deviations $\bar{\sigma}_{\lambda}$, the two right-side columns of table\,\ref{Tab:SigmaLambdRawResidual} indicate that the selected squared visibilities $V^2_{20\%}$ and the rescaled differential one have very comparable scattering along the wavelength dimension. 

Note that the rescaling factors $V^2_m \slash \bar{\mu} _{v^2}(i)$, applied on each frame to get the differential visibility from the absolute ones, are on average $>1$, which increases the scattering along wavelength. Therefore we would have $\bar{\sigma}_{\lambda}(\Delta V^2_{\rm R}) > \bar{\sigma}_{\lambda}(V^2)$ if we considered that quantity {\it per frame}. ( It would also be larger ; the computed $\bar{\sigma}_{\lambda}$ per frame, note presented in the table, are typically 2 to 3 times higher than the quantities in  the two right-side columns of table\,\ref{Tab:SigmaLambdRawResidual}). As we are here discussing the wavelength scattering computed from the averaged squared visibilities of the considered frames selection, the larger sample of $\Delta V^2_{\rm R}$ produces a smoother averaging effect than the $V^2_{20\%}$ sample. Eventually these two effect appear to balance and $\sigma_{\lambda}(\Delta V^2_{\rm R}) \approx \sigma_{\lambda}(V^2_{20\%})$ over an exposure file. 

This applies either on uncorrected or on piston-corrected squared visibilities, the latter quantity having standard deviations reduced by a factor 3 for datasets UT-MR-NoFT and AT-LR-FT, but only a marginal gain for the more scattered (and less piston-degraded) dataset AT-MR-FT, as illustrated in figure \,\ref{Fig:RmsAvgAT}.
\subsection{Best-fitting distribution laws}
\label{subsec:GofStats} 

\subsubsection{Method}

We present the results of $\chi^2$ Goodness-of-Fit (GoF) tests, which aim at determining whether the empirical distribution of a data sample is or not compatible with a standard distribution, 
with appropriately fitted parameters \citep{lehmann2005testing}. The binary outputs of the tests are obtained according to a predefined probability of false alarm ($P_{FA}$), fixed here at $5\%$. 

Below  is the list of the statistical laws which we tested and their associated parameters ($\mu$ is the location parameter, $\sigma$ and $\eta$ are scales parameters and $\nu$ stands for another possible shape parameter; the expressions of these distributions are recalled in Appendix \ref{lesdist}): 

\begin{itemize}
\item Normal (${\cal{N}}$), a function of $\mu$ and $\sigma$.
\item Student (t), a function of $\mu$, $\eta$ and $\nu_S$.
\item Log Normal (Log ${\cal{N}}$), a function of $\mu$ and $\sigma$.
\item Cauchy (${\cal{C}}$), a function of $\mu$ and $\nu_C$.
\end{itemize}

Let us first remark that even if  the squared visibilities are  normally distributed, this will not be the case for the differential ones. Indeed,  the definition of  $\Delta V^2$ in eq. (\ref{eq:DeltaV}) involves a ratio of  normal random variables. It is well known that this ratio  leads to a Cauchy  distribution \citep{Marsaglia} characterized by the parameters of the two normal distributions involved in the ratio. Between the normal  and the Cauchy distributions, the Student  distribution is able to fit both a Cauchy distribution ($\nu_S=1$) and a Normal distribution ($\nu_S  \to \infty$). 
We note finally that the Log Normal distribution was  used in order to fit the distribution of the squared visibilities by \cite{AMBER_Diff}. 
 
For the presented $\chi^2$ tests, the null hypothesis $\mathcal{H}_0$ is that the data is drawn from the tested distribution, and the alternative $\mathcal{H}_1$  that it is not. 
The binary result of a test is  $0$ if the  distribution tested under $\mathcal{H}_0$ is accepted, and $1$ otherwise.
  Obviously, even when a data sample is actually drawn from the distribution assumed under $\mathcal{H}_0$, there is always a possibility that the empirical distribution
 substantially deviates from the distribution under $\mathcal{H}_0$ because of estimation noise caused by the finite number of data samples and by the parameter fitting.

When the GoF test does not involve parameter fitting, results based on asymptotic theory allows to fix accurately the threshold corresponding  to the probability of 
false alarm ($P_{FA}$) \citep{lehmann2005testing}.  In Tables \ref{Tab:chi2GoFAll} and \ref{Tab:chi2GoFAllNopist} for instance, this threshold was set so that there
is $5 \%$ chances that data samples actually drawn from the null distribution are erroneously rejected by a GoF test without fit. This is verified by the second value
given in the ``Ctrl'' lines in the Tables. These values are  the rejection rate obtained for data drawn from the tested distribution and containing  the same number of  
samples as the tested interferometric data. 

When parameters are estimated in the GoF test  however, the former threshold guaranteeing a $P_{FA}$ of $5\%$ leads to a different rejection rate
\footnote{Since the number of points is limited, the distribution tested with estimated parameters leads to a better fitting power and thus a lower rejection rate than without an estimation}. 
Assessing analytically the relationship between $P_{FA}$ and test threshold is much more involved when parameters are estimated, but 
we can  resort to Monte-Carlo simulations to control the actual level of wrong rejections corresponding  to the $5\%$ threshold of the case without estimation. The
observed values are given by the first values of the ``Ctrl'' lines in the tables. 

\subsubsection{Goodness-of-fit results}

The results of statistical compatibility tests for the three datasets are presented in table\,\ref{Tab:chi2GoFAll}, and in table\,\ref{Tab:chi2GoFAllNopist} for the data corrected from piston and jitter chromatic effects.
The numbers represent the observed rejection rates of the null hypothesis and are expressed in percents. A lower value obtained for $V^2_{100\%}$, $V^2_{20\%}$ and $\Delta V^2_R$ means a higher compatibility with the laws reported in the corresponding  columns. The indicative size of the tests can be read in the first value of the ``Ctrl''  lines. 

For the three datasets, and either with or without a fitted correction of the piston effect, the Student distribution presents the lowest rate of rejection. 
In particular, the Student law is clearly favored against the normal distribution when considering the distribution along the wavelength dimension. Along the time dimension, except for the differential squared visibilities, the difference is less pronounced. { Even though its rejection rate is relatively lower, the Student law is logically rejected for absolute visibilities without a frame selection, for datasets showing a clear asymmetric distribution, either due to low visibility level (UT-MR-NoFT) or multimodal behaviour (AT-MR-FT).}

Finally, the Log Normal and Cauchy laws are most often associated to higher rejection rates than the other two tested distributions. 
The low-order correction of the piston effect does not modify qualitatively these results and similarly acts in favor of the Student distribution.

\begin{table}[ht!]  
\caption{Table of Goodness-of-Fit (GoF) rejection rate of the squared visibilities distributions along wavelengths (left-hand columns, results averaged on frames and files) and along time (right-hand columns, results averaged on wavelengths and files), with four standard distribution laws: Normal, Student, Log Normal and Cauchy. 
A low rate value means a higher compatibility with a given statistical law.
Numbers in italic font indicate that the result is not significant because the considered dataset and dimension contains too few points (here, only 13 spectral channels in AT-LR-FT data). See text for explanations on the "Ctrl" lines. 
} 
\label{Tab:chi2GoFAll}  
\vspace{-5mm}
\begin{center}
\small
 \begin{tabular}{cc|c|c|c|c|c|c|c|}  
 \cline{2-9} &  \multicolumn{4}{|c|}{Over wavelengths} & \multicolumn{4}{|c|}{Over time} \\ \cline{2-9}
 UT-MR-NoFT 
 &\multicolumn{1}{|c|}{ \bt ${\cal{N}}$ \et} 	& \bt  t \et	&  \bt  Log ${\cal{N}}$ \et &\multicolumn{1}{|c|}{ \bt ${\cal{C}}$ \et}
 &  \bt ${\cal{N}}$ \et 	& \bt t \et &  \bt  Log ${\cal{N}}$ \et&\multicolumn{1}{|c|}{ \bt ${\cal{C}}$ \et} \\ %
  \hline

 \multicolumn{1}{|c|}{$V^2_{100}$}                  	& 77 	& \textbf{55} 	& 85 	 &	99      & 99   & \textbf{69}    & 100 	&100     \\ \hline
 \multicolumn{1}{|c|}{Ctrl $V^2_{100}$}             & 3|5 & 2|4 & 2|5  &	4|5  	& 3|5	& 3|5	& 3|5	&3|4  \\ \hline
 \multicolumn{1}{|c|}{$V^2_{20}$}                   & 70 	& \textbf{57} 	& 81 	 &	99      & {15}    & \textbf{4}    & 34 	&99     \\ \hline
 \multicolumn{1}{|c|}{Ctrl $V^2_{20}$}              & 3|4 & 1|4 & 2|4  &	3|4  	& 2|4	& 2|5 & 2|5 &4|5  \\ \hline
 \multicolumn{1}{|c|}{$\Delta V^2_{\rm R}$} 		& 77 	& \textbf{45} 	& 84     &	99      & 67& \textbf{30}        & 100    &100     \\ \hline
 \multicolumn{1}{|c|}{Ctrl $\Delta V^2_{\rm R}$} 	& 2|5 & 2|5 & 2|5  &	3|5 	& 3|4	& 2|4	& 2|4 &3|4  \\ \hline
    AT-MR-FT\\   \hline
 \multicolumn{1}{|c|}{$V^2_{100}$}                  & 59 	& \textbf{29} 	& 53 	 &	100      & 34    & \textbf{11}    & 74 	&9     \\ \hline
 \multicolumn{1}{|c|}{Ctrl $V^2_{100}$}             & 3|5 & 1|4 & 2|5  &	4|5  	& 2|5	& 1|5	& 2|6	&2|6  \\ \hline
 \multicolumn{1}{|c|}{$V^2_{20}$}                   & 59 	& \textbf{22} 	& 49 	 &	100      & 1    & \textbf{0}    & 1 	&4     \\ \hline
 \multicolumn{1}{|c|}{Ctrl $V^2_{20}$}              & 2|5 & 0|4 & 3|5  &	3|4  	& 1|6	& 0|6 & 1|6 &1|7  \\ \hline
 \multicolumn{1}{|c|}{$\Delta V^2_{\rm R}$} 		& 59 	& \textbf{19} 	& 53     &	99      & 22    & \textbf{1}    & 29    &5     \\ \hline
 \multicolumn{1}{|c|}{Ctrl $\Delta V^2_{\rm R}$} 	& 2|4 & 1|5 & 2|5  &	4|5 	& 1|5	& 1|5	& 2|5 &2|6  \\ \hline
  AT-LR-FT\\   \hline
 \multicolumn{1}{|c|}{$V^2_{100}$}                  & \textit{23} 	& \textit{9} 	& \textit{19} 	 &	\textit{10}      & 95    & \textbf{84}    & 100 	&100     \\ \hline
 \multicolumn{1}{|c|}{Ctrl $V^2_{100}$}             & 1|6 & 0|6 & 1|6  &	1|6  	& 3|4	& 3|5	& 2|5	&4|5  \\ \hline
 \multicolumn{1}{|c|}{$V^2_{20}$}                   & \textit{12} 	& \textit{0} 	& \textit{10} 	 &	\textit{12}      & 5    & \textbf{4}    & 14 	&98     \\ \hline
 \multicolumn{1}{|c|}{Ctrl $V^2_{20}$}              & 1|6 & 0|6 & 1|6  &	1|7  	& 2|4	& 2|5 & 2|4 &4|6  \\ \hline
 \multicolumn{1}{|c|}{$\Delta V^2_{\rm R}$} 		& \textit{3} 	& \textit{0} 	& \textit{3}     &	\textit{10}      & 99    & \textbf{27}    & 99    &98     \\ \hline
 \multicolumn{1}{|c|}{Ctrl $\Delta V^2_{\rm R}$} 	& 1|6 & 0|5 & 1|6  &	1|7 	& 3|4	& 2|5	& 3|5 &3|4  \\ \hline
   \end{tabular}
\end{center} 
\end{table} 

\begin{table}[ht!]   \vspace{2mm}
\caption{Same as table\,\ref{Tab:chi2GoFAll}, here with squared visibilities corrected in each frame from the piston and jitter effects using a low-order polynomial fit along wavelength (absolute and differential $V^2_{\rm{nopist.}}$)} 
\label{Tab:chi2GoFAllNopist}  
\vspace{0mm}
\begin{center}
\small
 \begin{tabular}{cc|c|c|c|c|c|c|c|}  
 \cline{2-9} &  \multicolumn{4}{|c|}{Over wavelengths} & \multicolumn{4}{|c|}{Over time} \\ \cline{2-9}
 UT-MR-NoFT 
 &\multicolumn{1}{|c|}{ \bt ${\cal{N}}$ \et} 	& \bt  t \et	&  \bt  Log ${\cal{N}}$ \et &\multicolumn{1}{|c|}{ \bt ${\cal{C}}$ \et}
 &  \bt ${\cal{N}}$ \et 	& \bt t \et &  \bt  Log ${\cal{N}}$ \et&\multicolumn{1}{|c|}{ \bt ${\cal{C}}$ \et} \\ \hline
 \multicolumn{1}{|c|}{$V^2_{100}$}                  & 36 	& \textbf{18} 	& 43 	 &	99      & 85    & \textbf{75}    & 100 	&100     \\ \hline
 \multicolumn{1}{|c|}{Ctrl $V^2_{100}$}             & 2|4 & 2|5 & 2|5  &	3|5  	& 2|4	& 3|5	& 3|5	&4|5  \\ \hline
 \multicolumn{1}{|c|}{$V^2_{20}$}                   & 15 	& \textbf{10} 	& 22 	 &	99      & 14    & \textbf{7}    & 16 	&99     \\ \hline
 \multicolumn{1}{|c|}{Ctrl $V^2_{20}$}              & 3|5 & 2|5 & 3|5  &	3|5  	& 2|4	& 1|4 & 2|5 &3|5  \\ \hline
 \multicolumn{1}{|c|}{$\Delta V^2_{\rm R}$} 		& 37 	& \textbf{12} 	& 43     &	99      & 99    & \textbf{17}    & 98    &99     \\ \hline
 \multicolumn{1}{|c|}{Ctrl $\Delta V^2_{\rm R}$} 	& 2|5 & 2|4 & 2|5  &	4|6 	& 3|6	& 2|5	& 3|5 &3|4  \\ \hline
  AT-MR-FT\\    \hline
 \multicolumn{1}{|c|}{$V^2_{100}$}                  & 88 	& \textbf{12} 	& 51 	 &	99      & 34    & \textbf{11}    & 70 	&18     \\ \hline
 \multicolumn{1}{|c|}{Ctrl $V^2_{100}$}             & 2|4 & 1|4 & 2|4  &	4|5  	& 2|5	& 1|5	& 3|6	&2|6  \\ \hline
 \multicolumn{1}{|c|}{$V^2_{20}$}                   & 83 	& \textbf{5} 	& 52 	 &	100      & 11    & \textbf{0}    & 12 	&10     \\ \hline
 \multicolumn{1}{|c|}{Ctrl $V^2_{20}$}              & 2|5 & 2|6 & 2|5  &	5|8  	& 1|6	& 0|6 & 1|7 &1|7  \\ \hline
 \multicolumn{1}{|c|}{$\Delta V^2_{\rm R}$} 		& 88 	& \textbf{12} 	& 51     &	99      & 15    & \textbf{1}    & 16    &16     \\ \hline
 \multicolumn{1}{|c|}{Ctrl $\Delta V^2_{\rm R}$} 	& 2|5 & 1|4 & 2|5  &	4|5 	& 2|5	& 1|5	& 2|5 &2|6  \\ \hline
      AT-LR-FT\\   \hline
 \multicolumn{1}{|c|}{$V^2_{100}$}                  & \textit{1} 	& \textit{0} 	& \textit{2} 	 &	\textit{9}      & 96    & \textbf{71}    & 100 	&100     \\ \hline
 \multicolumn{1}{|c|}{Ctrl $V^2_{100}$}             & 1|6 & 0|6 & 1|6  &	1|7  	& 2|4	& 3|4	& 2|5	&5|6  \\ \hline
 \multicolumn{1}{|c|}{$V^2_{20}$}                   & \textit{1} 	& \textit{0} 	& \textit{1} 	 &	\textit{8}      & 14    & \textbf{4}    & 10 	&99     \\ \hline
 \multicolumn{1}{|c|}{Ctrl $V^2_{20}$}              & 1|5 & 0|6 & 1|6  &	1|6  	& 2|5	& 2|5 & 2|4 &3|5  \\ \hline
 \multicolumn{1}{|c|}{$\Delta V^2_{\rm R}$} 		& \textit{2} 	& \textit{0} 	& \textit{2}     &	\textit{9}      & 98    & \textbf{9}    & 96    &100     \\ \hline
 \multicolumn{1}{|c|}{Ctrl $\Delta V^2_{\rm R}$} 	& 1|6 & 0|6 & 1|6  &	1|7 	& 3|4	& 2|5	& 3|4 &4|5  \\ \hline
   \end{tabular} 
\end{center}
\end{table}

These results come in contrast to the rather generally accepted idea that squared visibilities are normally or log-normally distributed \citep{FlorentinDataAmberCalibration}. 
 The estimator used to provide the AMBER squared visibilities is actually expressed as a ratio of random variables \citep{amdlib2007}. This is
 probably the reason why the Student law, which is a flexible ratio distribution, gives the best fit. 

\begin{figure*}[ht!]
\begin{center}
\includegraphics[width=\textwidth]{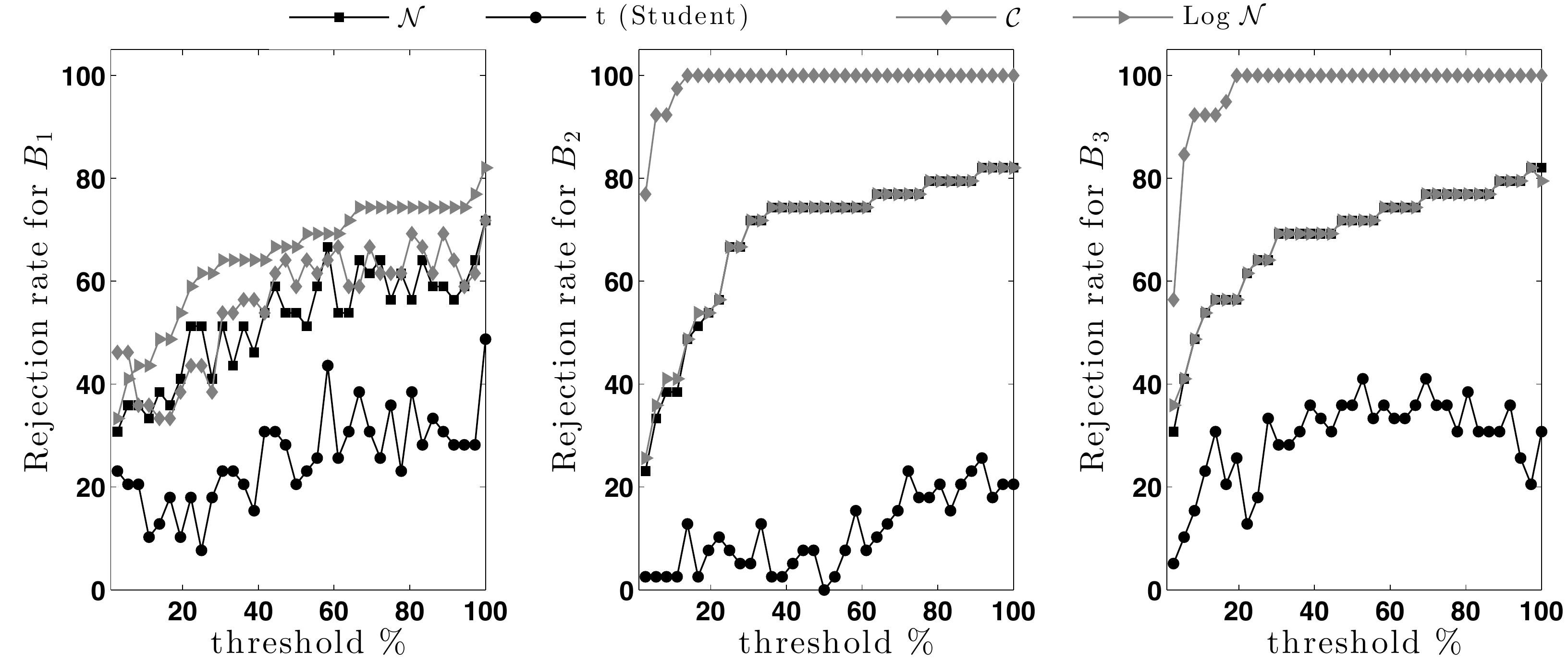}
\caption{$\chi^2$ GoF test applied on average visibilities as a function of the frame selection threshold applied on the AT-MR-FT data set. The rejection rate is in \% and is averaged over exposure files, for each baseline.}
\label{Fig:GofMean}
\end{center}
\end{figure*}

The same analysis was performed on {\it temporally averaged visibilities} in order to characterize their chromatic distribution, as a function of "best SNR" frame percentage. Here, visibilities are averaged over the frames selection set of each exposure file, which corresponds to the way most Amber users would get their reduced data. 
Figure\,\ref{Fig:GofMean} represents the GoF rejection rate (in percents) for each threshold value and base, from AT-MR-FT dataset. In accordance with previous non-averaged results, it clearly shows that the Student distribution presents to best fit of averaged data distribution, regardless of the selection threshold. 

\label{subsec:GofMean} 

\subsection{Correlations between squared visibilities}

How, and how much the squared visibilities from different wavelengths are correlated with each other is an important issue for tackling the polychromatic aspect of the inverse problem. 
Indeed, covariances matrices arise naturally through the likelihood for instance \citep{MIRA,meimon2005}. But the absence of measures leads in practice to neglect any 
kind of dependence between the observables. 

Between two spectral channels (say, $n$ and $p$) and for a given dataset and baseline, the coefficients of the correlation matrix are: 

\begin{eqnarray}
	\mathrm{c}\,(n,p)= \frac{ \avg{ \left[ V^2_n - \avg{V^2_n }_t \right] \left[V^2_p - \avg{V^2_p}_t\right] }_t } {\sigma(V^2_n) ~\sigma(V^2_p)}.
\label{eq:corrij}
\end{eqnarray}

\begin{figure*}[ht!]
\begin{minipage}[l]{.45\linewidth}
\begin{center}	
\includegraphics[height=.95\textwidth]{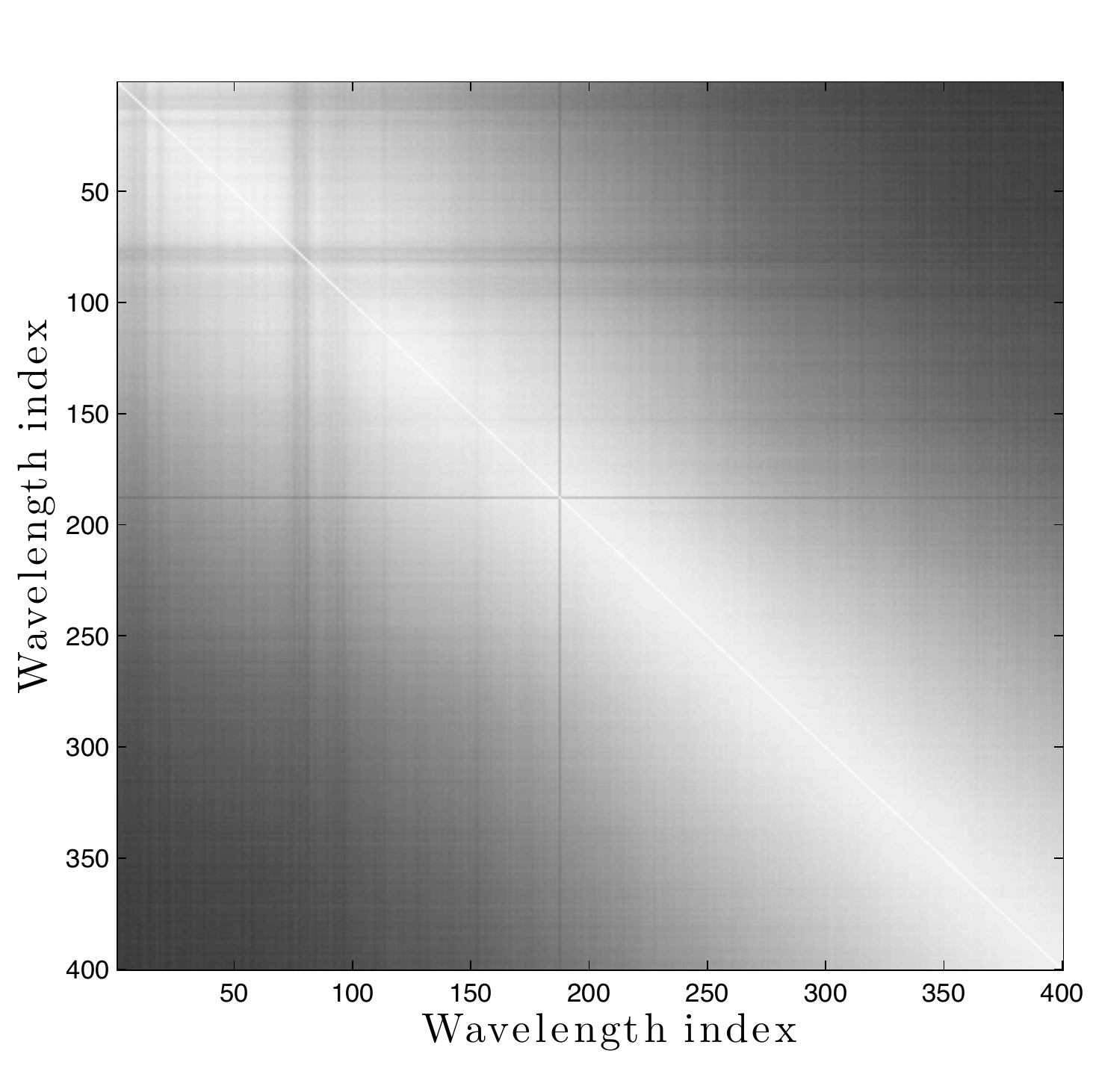}
\end{center}
\end{minipage}
\begin{minipage}[r]{.45\linewidth}
\begin{center}
\includegraphics[height=.95\textwidth]{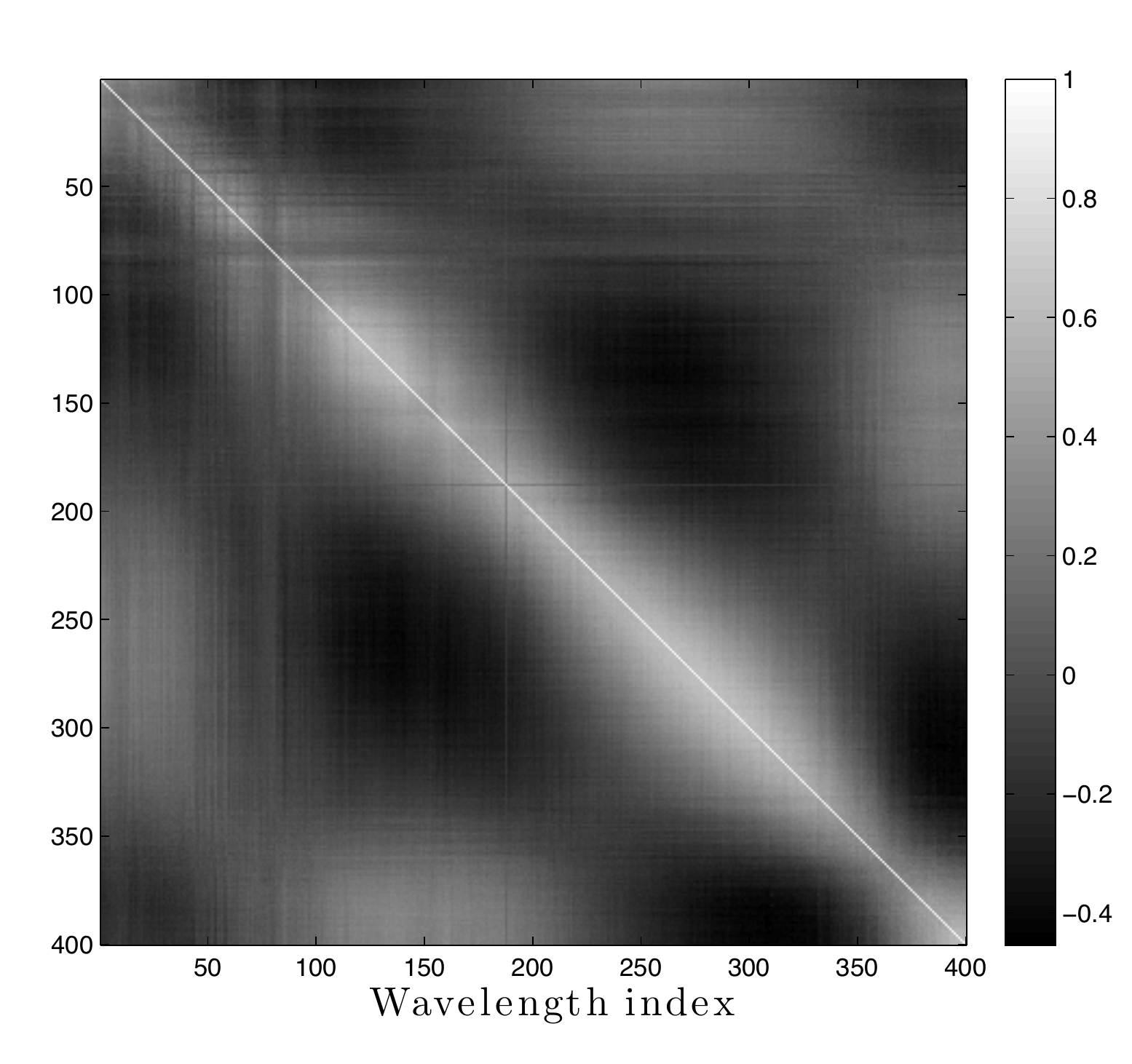}
\end{center}
\end{minipage}

\caption{Examples of matrices of the temporal correlation coefficients between spectral channels (see eq.\,\ref{eq:corrij}) for dataset UT-MT-NoFT, averaged over observation files and over the three baselines. \emph{Left:} $V^2_{20\%}$, no correction of the piston effect, \emph{Right:} $\Delta V^2_{\rm R}$, after piston-fit correction. The value of each correlation coefficient is proportional to the gray scale \emph{(far right)}.}
\label{Fig:MatricesCorr}
\end{figure*}

This quantity  also describes  the entries of the sample covariance matrix normalized by the product of the standard deviations $\sigma(V^2_n)$ and $\sigma(V^2_p)$ (therefore $\mathrm{c}\,(n,p)=1$ for $n=p$) as illustrated in figure \,\ref{Fig:MatricesCorr}.
 
\begin{figure}[ht!]
\begin{center}
\includegraphics[width=.5\textwidth]{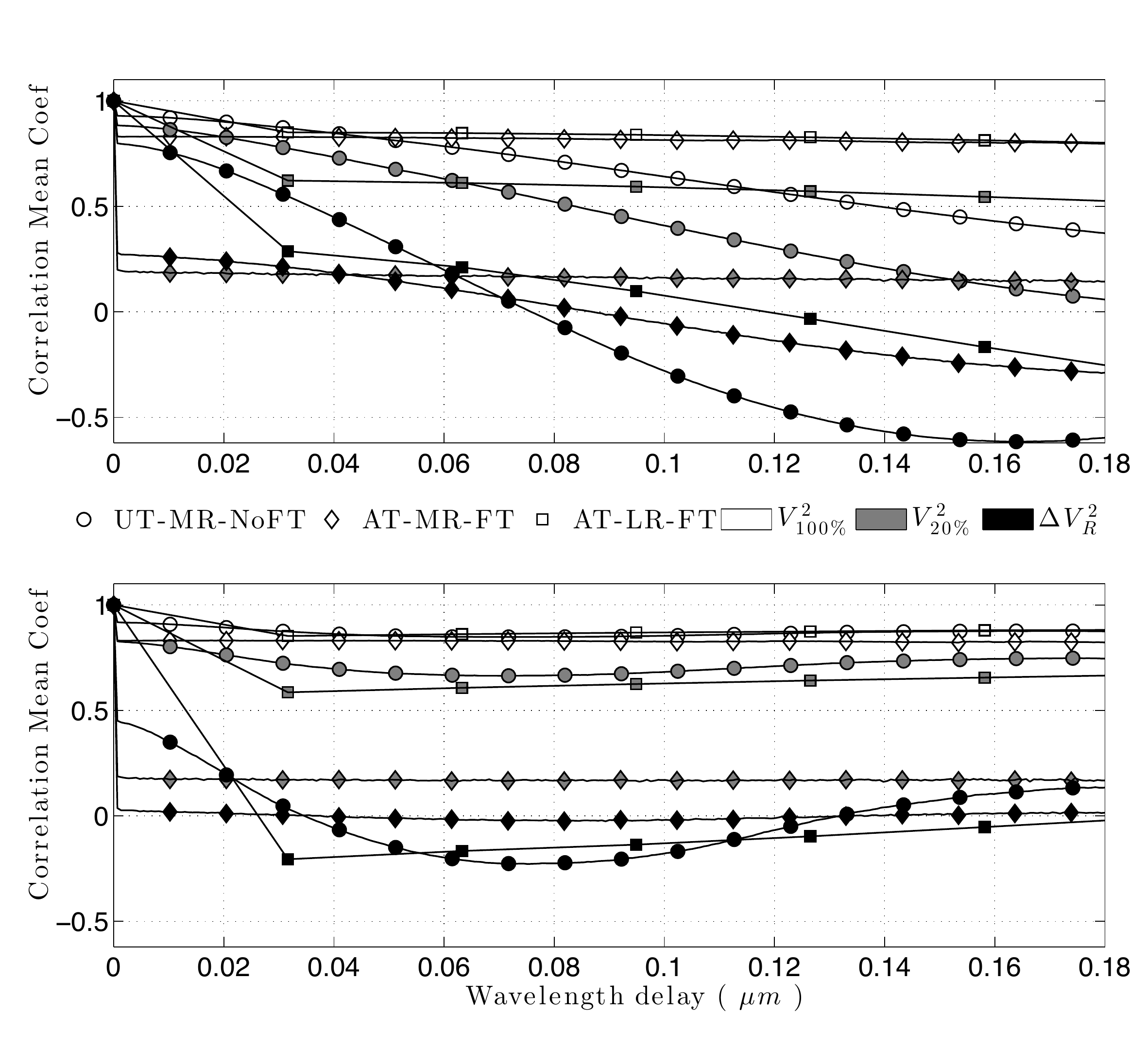}
\caption{Coefficients of instantaneous correlation between spectral channels, averaged over observation files, for the three datasets as a function of the wavelength separation $k$, expressed in microns. Each curve is issued from a correlations matrix after averaging eq.\,\ref{eq:corrk}. Datasets UT-MR-NoFT, AT-MR-FT and AT-LR-FT are represented respectively by circles, diamonds and square symbols (for clarity purposes, only one of each 15 correlation points of the two medium-resolution datasets are shown). The squared visibilities with 100\% and 20\% selection thresholds and the rescaled differential squared visibility appear respectively as non-filled (white), gray and black symbols. \emph{Top}: without correction of the piston effect, \emph{Bottom:} after frame-by-frame subtraction of the fitted piston effect.}
\label{Fig:AutoCorr}
\end{center}
\end{figure}

The coefficients defined by eq.\,(\ref{eq:corrij}) may be shown as "correlation images" of dimensions $[n_\lambda \times n_\lambda]$, such as the examples in figure \,\ref{Fig:MatricesCorr}. Nevertheless, these results are, on one hand, very much space-consuming (as we want to compare correlations for the different squared visibility estimators $V^2_{20\%}$, $V^2_{100\%}$ and $\Delta {V^2_{\rm R}}$, datasets and "piston-correction" options), and on the other hand $\mathrm{c}\,(n,p)$ appears to depend on the wavelength index difference $k=n-p$ (with $n>p$) rather than on the considered channel $p$ itself. 
The average instantaneous correlations  as a function of the spectral separation $k$ 
 are shown if figure \,\ref{Fig:AutoCorr} for all the squared visibilities, datasets, and with and without correction for piston. 

\begin{eqnarray}
\rho(k) = \frac{1}{n_\lambda -k} \sum_{n=1}^{n \le n_\lambda -k} \mathrm{c}\,(n,n+k) 
\label{eq:corrk}
\end{eqnarray}

The  squared visibilities without selection ($V^2_{100\%}$)  are in all cases very much correlated between wavelengths: this is explained by considering that the sudden visibility losses affect essentially all  wavelengths simultaneously and in a similar way. 

This effect is very much reduced when considering the selected squared visibilities ($V^2_{20\%}$). The reason is that on average these frames with a higher SNR  are filtered and undergo smaller perturbations (global visibility losses and large pistons). Hence, the correlating effect caused by erratic visibility losses is less pregnant in these frames. 
It is not the case for the two other data set and we conclude that at a selection rate of $20\%$, some selected frames of UT-MR-NoFT and AT-LR-FT data sets still suffer from visibility loss and piston.

Interestingly, sudden visibility losses tend to instantaneously correlate the squared visibilities at all wavelengths.
But piston (with slopes that are either positive or negative along the frames) tends to decorrelate them.
This is why when piston is suppressed (bottom figures), the correlation of visibilities increase.


Turning now to the  rescaled differential squared visibilities $\Delta V^2_{\rm R}$, their correlations present a different behaviour. 
For the $\Delta V^2_{\rm R}$ the impact of visibility losses between frames is removed
 by construction, which  globally reduces the correlations. But without frame selection, the piston effect remains entirely, and piston is actually a correlating effect for these quantities. Indeed,
 whatever the slope of the piston, differential squared visibilities at both ends of the spectrum tend to be opposite
side w.r.t. the average level of the frame. This induces an anticorrelation of differential squared visibilities at large $k$. Conversely, differential squared visibilities at near wavelengths tend to be on the same side of the average level, and thus correlated.
When correction for piston is made for $\Delta V^2_{\rm R}$, this effect is logically  reduced.

We may now investigate the cross-correlations $\mathrm{c}\,(B_r,B_s)$ 
between the squared visibilities from different baselines $B_r$ and $B_s$, with $r \ne s$. Those are analyzed without considering any shift between the spectral channels (i.e. $k=0$). $\mathrm{c}\,(B_r,B_s)$  is then computed similarly as in eq.\,\ref{eq:corrij}, except that $r$ and $s$ refer now to baseline indices, and that the averaging operator applies only on frames measured simultaneously on the baselines.
It is difficult to analyze the instantaneous cross correlation between bases for the $V^2_{20\%}$ because this requires to have simultaneously selected frames (which is usually not the case at a selection rate of $20\%$).

The results are summarized in Table\ref{Tab:InterCorr}. First of all, it appears that the correlation depends largely on the considered pair of baseline, and that two pairs (here, $B_1 ~ B_3$ and $B_2 ~ B_3$, for which $\mathrm{c}\,>0.2$) are much (more) correlated than the third one ($B_1 ~ B_2$, where $\mathrm{c}\, \le 0.18$).
This might translate the fact that some parts of the interferometric chain produce time-variable effects (e.g. a suboptimal adaptive optics or a vibrating telescope, turbulences from a longer delay line path, non-centered beam injection within the optical fiber\dots which impact at least one pair of baselines simultaneously, through visibility losses and possibly through a slope of squared visibilities vs wavelengths. 

As for how correlations vary with the considered visibility estimator, it appears that for  squared visibilities $V^2_{100\%}$, the piston-fit subtraction has virtually no effect. But its differential counterpart has a lower correlation  (40\% to 60\% for the non-piston corrected case, and even lower in the case of piston-fit corrected data.) 

\begin{table}[ht!]   \vspace{0mm}
\caption{Coefficients of the instantaneous cross-correlation between pairs of baselines, for unselected squared visibilities ($V^2_{100\% }$) and rescaled differential ones ($ \Delta V^2_{\rm R}$). } 
\vspace{5mm}
\begin{tabular}{cc|c|c|c|}
\cline{3-5}
\bt \et &\bt Baselines \et  & \bt UT-MR-NoFT \et & \bt AT-MR-FT \et & \bt AT-LR-FT \et \\ \hline

 \multicolumn{1}{|c}{\multirow{3}{*}{$V^2_{100\%}$}} & 
 \multicolumn{1}{|c|}{$B_1$ $B_2$} & 0.18  & 0.098  & 0.13     \\ \cline{2-5} 
 \multicolumn{1}{|c}{}                        & 
 \multicolumn{1}{|c|}{$B_1$ $B_3$} & 0.23 & 0.31 & 0.47     \\ \cline{2-5} 
 \multicolumn{1}{|c}{}                        & 
 \multicolumn{1}{|c|}{$B_2$ $B_3$} & 0.28 & 0.79 & 0.40     \\ \cline{1-5} 
 \multicolumn{1}{|c}{\multirow{3}{*}{$\Delta V^2_{R}$}} & 
 \multicolumn{1}{|c|}{$B_1$ $B_2$} & 0.12 & 0.004 & 0.095     \\ \cline{2-5} 
 \multicolumn{1}{|c}{}                        & 
 \multicolumn{1}{|c|}{$B_1$ $B_3$} & 0.13 & 0.15 & 0.22     \\ \cline{2-5} 
 \multicolumn{1}{|c}{}                        & 
 \multicolumn{1}{|c|}{$B_2$ $B_3$} & 0.14 & 0.37 & 0.17     \\ \cline{1-5} \\ \hline 
 \multicolumn{1}{|c}{\multirow{3}{*}{$V^2_{100, \rm{nopist.}}$}} & 
 \multicolumn{1}{|c|}{$B_1$ $B_2$} & 0.15 & 0.15 & 0.13     \\ \cline{2-5} 
 \multicolumn{1}{|c}{}                        & 
 \multicolumn{1}{|c|}{$B_1$ $B_3$} & 0.20 & 0.34 & 0.47     \\ \cline{2-5} 
 \multicolumn{1}{|c}{}                        & 
 \multicolumn{1}{|c|}{$B_2$ $B_3$} & 0.26 & 0.80 & 0.44     \\ \cline{1-5} 
 \multicolumn{1}{|c}{\multirow{3}{*}{$\Delta V^2_{\rm R, nopist.}$}} & 
 \multicolumn{1}{|c|}{$B_1$ $B_2$} & 0.08 & 0.09 & 0.08     \\ \cline{2-5} 
 \multicolumn{1}{|c}{}                        & 
 \multicolumn{1}{|c|}{$B_1$ $B_3$} & 0.09 & 0.19 & 0.19     \\ \cline{2-5} 
 \multicolumn{1}{|c}{}                        & 
 \multicolumn{1}{|c|}{$B_2$ $B_3$} & 0.10 & 0.32 & 0.12     \\ \cline{1-5}
\end{tabular}
\label{Tab:InterCorr}  
\end{table} \vspace{0mm}


\section{Conclusions and perspectives}

This paper provides a detailed statistical analysis framework for interferometric squared visibilities through the example of AMBER's data.
Several conclusions arise from this study. 

Regarding squared visibilities, we could see that devising an automatic procedure for optimum threshold selection is difficult.  A predetermined SNR selection value might result,
depending on the observing conditions, in either a strongly biased sample of frames (whose distribution contains secondary modes, left-wing asymmetry and/or over-represented tails), or in a severe selection which increases unnecessarily the variance of estimation. We have however provided ideas which should be worked out
based on the stability of the observed  distributions of the frame and on their modes.  

Colour-differential squared visibility appears, on the other hand, as a more stable and regular quantity. 
When rescaled to an average level estimated from a "best frames" sample of the squared visibility, its statistical standard deviation over time is typically improved by 25\% with respect to the usual squared visibility estimator, whereas it behaves very similarly along the spectral dimension. This result is due to the fact that differential squared visibilities allow to take into account a significantly larger fraction of the data, and thus benefit from a reduced temporal dispersion. Because of the centering of the rescaled differential squared visibilities of each frame around a common average value, their distribution is also more clearly characterized by the Goodness-of-Fit test than their absolute counterparts. Also, re-scaled differential squared visibilities have a lower spectro-temporal correlation than the absolute ones 
Although these results depend very much on the dataset (with different instrumental setups and ambient conditions), they nevertheless indicate that re-scaled differential squared visibilities usually constitute a valuable alternative estimator of squared visibility.

Regarding the statistical law that best describe both $V^2$ and $\Delta V^2_{\rm R}$, we find a better fit for the Student distribution than for others, in particular than the normal distribution. 
{ The impact of assuming one or the other of these statistics for model-fitting has been investigated in a preliminary work \citep{Schutz2013}, using our AT-MR-FT observations for generating semi-synthetic data (visibilities of a synthetic uniform disk, to which was added real noise derived from the observations). With actual diameters ranging from 0.1 to 20 mas, the accuracy on the estimated diameter is improved (by a factor up to $\approx 2$) by introducing a Student instead of Gaussian likelihood. This study should be extended to other data sets and sources models in order to assess this effect in a more general context. 
We finally note that in \cite{lange89} and, more recently, \cite{Kazemi} the Student distribution was also used to improve the parameter estimation in model-fitting.
}

We finally find clear correlation effects caused by atmospheric perturbations. Accounting for such correlations should indeed improve the models used in inverse problems and thus
the accuracy of model fitting and image reconstruction results. 

{ We expect that these results should equally apply to other existing data reduction softwares using either amdlib as their core, or based on similar ABCD-algorithm, as they would present the same principles and theoretical biases than the official amdlib DRS we used here. On the other hand, a DRS based on Fourier analysis might yield different results, especially on low visibility or poor SNR data. Although such prototype software has already given some significant astrophysical results on Amber data \citep{DiffVis2012}, it is currently not enough complete and robust for allowing a numerical comparison on the different cases studied in the present work.}

We are planning to extend our analysis to interferometric data obtained from instruments PIONIER \citep{PIONIER2011} and VEGA \citep{Vega2009}. Finally, a similar statistical study of the closure and differential phases is also under investigation.

\appendix

\section{Aberrant point removing in differential squared visibilities computation} \label{ann:diffvis}

Differential squared visibility histograms, without any other selection that the bad flags filtering included in the original data, show that points with very large or low values with respect to the average of 1 may be over-represented compared to the expected ``tail'' from an usual statistical distribution law, which does not allow to get a realistic fit using such standard law. The odd statistical points can be caused by a number of observational events (cosmic ray, bad pixel in CCD,\dots), not always corrected or flagged by the data reduction process, and not associated to an odd value of a selection criterion (such as the fringe SNR). Thus, they cannot be removed \textit{a priori}, and we chose to flag and remove them using their own squared visibility value, through an iterative process: differential squared visibilities whose difference with the average (1) exceeds a given threshold (say, 10 times the observed standard deviation $\sigma$) are identified by their frame and wavelength number, and discarded before computing a new set of differential squared visibilities, with its own standard deviation, etc\dots. A stable selection map, which contains more than 95\% of the  original data, is obtained over just a few of these iterations ; the rejection is therefore quantitatively marginal but proved to have significant effects when it comes to fitting a standard distribution law.  Such filter maps are used in the analysis proposed in this paper, not only for differential squared visibilities, but also for absolute squared visibilities, as a complement to the standard selection process.

\section{Distributions definition}
\label{lesdist}
The Normal law:
\begin{eqnarray*}
f(x;\mu,\sigma) = \frac{1}{\sigma \sqrt{2\pi}} \mathrm{exp}\left[{-\frac{1}{2}\left(\frac{x-\mu}{\sigma}\right)^2}\right]
\label{eq:NormalLaw} \end{eqnarray*} 

The Student law:
\begin{eqnarray*}
f(x;\mu,\sigma,\nu) = \frac{\Gamma((\nu+1) \slash{2})} {\sqrt{\pi\nu\eta^2 }\,\Gamma({\nu}\slash{2})} \left(1+\frac{\left(x-\mu \right)^2}{\nu\eta^2} \right)^{-(\nu+1)\slash{2}}
\label{eq:StudentLaw} \end{eqnarray*} 

The Log Normal law:
\begin{eqnarray*}
f(x;\mu,\sigma) = \frac{1}{x \sigma \sqrt{2 \pi}}\, \mathrm{exp}\left[{-\frac{(\ln x - \mu)^2}{2\sigma^2}}\right],\ \ x>0
\label{eq:LogNormal} \end{eqnarray*} 

The Cauchy law:
\begin{eqnarray*}
f(x; \mu,\nu) = \frac{1}{\pi\nu \left[1 + \left(({x - \mu})\slash{\nu}\right)^2\right]} = { 1 \over \pi } \left[ { \nu \over (x - \mu)^2 + \nu^2  } \right]
\label{eq:LogNormal} \end{eqnarray*}


\end{document}